# Elemental partitioning and isotopic fractionation of Zn between metal and silicate and geochemical estimation of the S content of the Earth's core

Brandon Mahan [a,*], Julien Siebert [a], Emily A. Pringle [a], Frédéric Moynier [a,b]

[a] *Institut de Physique du Globe de Paris, Université Paris Diderot, Sorbonne Paris Cité, CNRS UMR 7154, 1 rue Jussieu, 75238 Paris Cedex 05*
[b] *Institut Universitaire de France, Paris, France*
*corresponding author, email: mahan@ipgp.fr, phone: +33 (07) 62 93 56 12

## Abstract

Zinc metal-silicate fractionation provides experimental access to the conditions of core formation and Zn has been used to estimate the S contents of the Earth's core and of the bulk Earth, assuming that they share similar volatility and that Zn was not partitioned into the Earth's core. Therefore, Zn provides both direct and indirect information into the origin and eventual fate of volatile and siderophile elements on Earth. However, the partitioning of Zn between metal and silicate - as well as the associated isotopic fractionation - is not well known. We have conducted a suite of partitioning experiments to characterize Zn elemental partitioning and isotopic fractionation between metal and silicate as a function of time, temperature, and composition. Experiments were conducted at 2 GPa and temperatures from 1473K to 2273K in a piston cylinder apparatus, with run durations from 5 to 240 minutes for four distinct starting materials. Chemical and isotopic equilibrium is achieved within 10 minutes of experimental outset. Zinc metal-silicate isotopic fractionation displays no resolvable dependence on temperature, composition, or oxygen fugacity within the data set. Therefore, the Zn isotopic composition of silicate phases can be used as a proxy for bulk telluric bodies. Partitioning results from this study and data from literature were used to robustly parameterize Zn metal-silicate partitioning as a function of temperature, pressure, and redox state. Using this parametric characterization and viable formation conditions, we have estimated a range of Zn contents in the cores of iron meteorite parent bodies (i.e. iron meteorites) of ~0.1-150ppm, in good agreement with natural observations. We have also calculated the first geochemical estimates for the Zn contents of the Earth's core and of the bulk Earth, at 242 ±107ppm and 114 ±34ppm (respectively), that consider the slightly siderophile behavior of Zn. These estimates of the Zn contents of the Earth's core and bulk Earth are significantly higher than previous estimates 0-30ppm and 24-47ppm, respectively. Assuming similar volatility for S and Zn, a chondritic S/Zn ratio, and considering our new estimates, we have calculated a geochemical upper bound for the S content of the Earth's core of 6.3 ±1.9wt%. This indicates that S may be a major contributor to the density deficit of the Earth's core or that the S/Zn ratio for the Earth is non-chondritic.

**1. Introduction**

Earth's formation and subsequent differentiation is thought to have occurred by the accretion of large planetesimals (Kleine et al., 2002; Chambers, 2004; Wood et al., 2006). The impact energy of these coalescing planetary embryos combined with the radioactive decay of short-lived radioactive elements melted a substantial fraction of the proto-Earth (possibly multiple times), creating a deep magma ocean (up to 1100km depth) (Tonks and Melosh, 1993; Li and Agee, 1996; Chabot et al., 2005; Wade and Wood, 2005; Halliday and Wood, 2009). The canonical view of core formation holds that during this melting event, gravitational separation of immiscible silicate and metallic melts occurred along with the isotopic fractionation of constituent elements at high temperature and high pressure (HT-HP) between metal and silicate (core and mantle) phases (Wood et al., 2006; Georg et al., 2007; Shahar et al., 2011; Hin et al., 2013). In this scenario, metallic droplets form in the deep magma ocean in chemical and isotopic equilibrium and sink to the bottom under their own weight. This metallic downpour accumulates at the base of the magma ocean (the last point of metal-silicate equilibration), subsequently coalescing via cohesive forces to form large, inverted diapers that then rapidly descend to the continuously growing core without any further equilibration, making core formation the single largest chemical fractionation event in Earth's history (e.g. Li and Agee, 1996; Siebert et al., 2011). Although this generic scenario is well accepted, the physical conditions at which the process happened and therefore the composition of the Earth's core (especially with regard to residing light elements), as well as the origin and fate of its volatile elements, are still highly debated (Dreibus and Palme, 1996; Li and Agee, 1996; Wade and Wood, 2005; Corgne et al., 2008; Albarède, 2009; Brenan and Mcdonough, 2009; Siebert et al., 2011; Hin et al., 2013; Savage et al., 2015).

Zinc is a moderately volatile element with a 50% condensation temperature ($T_{50}$) of 726K (Lodders, 2003). It is comprised of five stable isotopes – $^{64}$Zn (48.6%), $^{66}$Zn (27.9%), $^{67}$Zn (4.1%), $^{68}$Zn (18.8%) and $^{70}$Zn (0.6%). The Zn isotopic composition is usually given as the per mil (‰) deviation of the $^{66}$Zn/$^{64}$Zn ratio from the JMC-Lyon standard ($\delta^{66}$Zn). The abundance of Zn in most planetary materials is high enough, even in the most volatile depleted samples, for precise elemental abundance and isotopic measurements in meteorites, lunar, and terrestrial samples from Earth; therefore Zn elemental and isotopic systematics have been used as a gauge of volatile depletion and a tracer of evaporation in planetary materials (e.g. Day and Moynier, 2014). Additionally, Zn has only one dominant oxidation state (2+) in the silicate fraction of planets, which limits the source of isotopic fractionation and renders Zn partitioning largely insensitive to the composition of the silicate melt (Siebert et al., 2011), thereby making interpretations of both Zn elemental partitioning and stable isotope data more straightforward. Therefore, Zn is very useful for understanding various planetary processes (evaporative loss, planetary differentiation) through direct study and also provides indirect insight into the behavior of similarly volatile elements, which are often more complicated and technically challenging to investigate independently. Sulfur ($T_{50}$ of 664K) is a possible - though largely debated - light element in the Earth's core (e.g. Badro et al., 2007; Morard et al., 2013). One of the first estimate of the S content of Earth's core was derived via its relative abundance to Zn in meteorites (Dreibus and Palme, 1996), wherein estimates of the Zn and S contents of the Earth's mantle and a chondritic S/Zn ratio were used, along with the assumption that Zn is completely lithophile, to calculate the S content of the core.

In order to study the Zn isotopic composition of a differentiated planetary body through surface rocks, it has been assumed that the Zn isotopic composition represents the bulk planetary body. Chen et al. (2013b) have shown that igneous differentiation (on Earth) could not cause large Zn isotopic fractionation (<0.1‰) and assumed that core formation would have a negligible isotopic effect as well. Chen et al. (2013b) considered that no Zn entered the Earth's core, however Zn may be slightly siderophile, especially at high temperature (Siebert et al., 2011), and therefore a fraction of the Zn budget of a differentiated planetary body could reside in its core. Bridgestock et al. (2014) investigated the effect of metal-silicate partitioning on Zn isotope fractionation. From three metal-silicate partitioning experiments (high purity oxide and metal powder starting materials; 1.5 GPa and 1650K) in MgO capsules, no resolvable Zn isotope fractionation between metal and silicate phases was found within a 0.1‰ range, and an estimated Zn metal-silicate partition coefficient of $D_{met-sil}^{Zn}$ ~0.7 was found for the experiments. This interesting first look into Zn metal-silicate isotopic fractionation invites further investigation, as these observations are at static conditions and therefore do not remark on the sensitivity of Zn metal-silicate isotopic fractionation to temperature or composition, and likewise the sensitivity of Zn metal-silicate partitioning to temperature, pressure, $fO_2$, and composition.

Several studies (e.g. Siebert et al., 2011; Bridgestock et al., 2014; Wood et al., 2014) have indicated that Zn in fact displays a slight affinity for the metallic phase (slightly siderophile behavior) in metal-silicate partitioning experiments, and each have contributed facets of understanding to the overall systematics of Zn metal-silicate partitioning in regards to temperature, pressure, $fO_2$, and the influence of other elements (e.g. carbon and sulfur). However, a more holistic and systematic approach is necessary to provide a better understanding of Zn metal-silicate partitioning and isotopic fractionation as a function of these varying conditions.

Here we present a suite of experiments to simultaneously determine Zn metal-silicate elemental partitioning and isotopic fractionation as a function of multiple influential parameters. By coupling experimental petrology and stable isotope geochemistry, this work explores the minimum chemical and isotopic equilibration time for metal-silicate differentiation (at 1673K) through an experimental time series, as well as the temperature dependence of Zn metal-silicate partitioning and isotopic fractionation across a range of temperatures from 1473K-2273K. The effect of S on partitioning and isotopic fractionation has been investigated through a comparison of S-bearing and non S-bearing experimental series, as well as the effect of Sn through a comparison of Sn-bearing and non Sn-bearing experimental series. The effect of silicate composition on isotopic fractionation, namely Si content, has been investigated through a series of experiments with a haplogranitic silicate component in comparison to experimental series with basaltic silicate components. The effects and efficacy of differing capsule materials has also been investigated through the comparison of complimentary experiments in BN (boron nitride) capsules with those in MgO.

The results of these experiments and subsequent parameterization of Zn metal-silicate partitioning using our data with available Zn partitioning data in the literature have been used to provide the first robust geochemical estimates of the Zn and S contents of the bulk Earth and Earth's core that take into account the slightly siderophile behavior of Zn and its sensitivity to the temperature, pressure, and fO$_2$ conditions of Earth's core formation and differentiation.

## 2. Petrologic and Chemical Methodology
### 2.1. Starting Materials

Four compositionally distinct starting materials (see Table 1) were used to assess controls on the elemental partitioning and isotopic fractionation of Zn. Three starting materials were made containing 60wt% mid-ocean ridge basalt (MORB), to which high-purity metallic or metal-bearing powders were added – [20wt% Fe, 20wt% FeS], [40 wt% Fe], and [24wt%Fe, 16wt% Sn], respectively. The high S content (20wt% FeS) in the metallic phase for the Fe-FeS starting material was chosen due to the efficacy of S to suppress the melting temperature of the Fe-rich metallic phase, enabling complete melting at low(er) temperatures (~1573-1973K) (Buono and Walker, 2011). Tin was used as an alternative melting point suppressor in the Fe-Sn experiments (Okamoto, 1993; Hin et al., 2013) for comparison to other starting materials, namely S-bearing experiments at lower temperature, and to explore any possible effects of Sn on Zn metal-silicate partitioning and isotopic fractionation. One starting material was made containing 60wt% haplogranite synthesized from high purity oxides, to which 20wt% Fe and 20wt% FeS were added. The synthetic haplogranite (from high purity oxides, with the same additional proportions of Fe, FeS) (Tuttle and Bowen, 1958) was used in order to achieve lower melting temperatures for the silicate phase, as well as to provide additional insight into the effect of composition, in this case Si content.

All starting materials were then doped with 1wt% Zn powder (Alfa Aesar, -100 mesh, 99.9% metals basis) to ensure robust measurement of Zn isotopic signatures and concentrations in both metal and silicate phases, and to overprint the Zn isotopic signature of the MORB in the starting materials. Each starting material was dry-mixed in an agate mortar and pestle to ensure a compositionally homogeneous mixture.

### 2.2. Experimentation

Experiments were conducted in a piston-cylinder apparatus at the Institut de Physique du Globe de Paris (IPGP). All experiments were conducted at 2 GPa, over a range of temperatures in a ½" cell assembly (Table 2). Experiments were heated with a cylindrical graphite furnace and temperature was measured using a top-loaded, Type D W/Re thermocouple. Lower temperature experimental cells contain concentric cylinders of Talc (outer) and Pyrex (inner) as the pressure media; higher temperature experiments (metallic Fe experiments) contain a single cylinder of barium carbonate ($BaCO_3$) pressure medium, chosen for its stability and superior thermal efficiency. Dehydrated MgO was used for internal spacers as well as sample capsules for the majority of experiments. Two additional experimental capsules were machined from boron nitride (BN). All capsules were cored concentrically to accommodate approximately 100mg of sample material, baked to volatilize water and any organic contaminants from machining, and filled manually prior to experimental runs. Sample capsules were centrally located in the cell to minimize thermal gradients, and a thin (~1.4mm) MgO/BN lid was placed atop the capsule to insulate it from thermocouple contamination while providing an accurate temperature measurement.

Compositions of all starting materials can be found in Table 1. A series of Fe-FeS + MORB experiments was conducted at 1673K over a range of durations from 5-240 minutes to constrain the timing of chemical and isotopic equilibrium. Two additional experiments with the same starting composition were conducted at 1823K and 1973K (15 minute and 3 minute run durations, respectively) to assess the effects of elevated temperature on Zn metal-silicate fractionation. A series of three experiments with a nominally S-free, Fe metallic phase (+ MORB) was conducted over a temperature range of 1973-2273K to determine an upper limit of analytically resolvable Zn fractionation and to investigate the influence of S, if any, on metal-silicate fractionation of Zn. Run durations for the experiments were varied depending on temperature (after Siebert et al., 2011), with progressively shorter durations at increased temperatures to mitigate capsule interaction and preferential loss of isotopically light Zn to the MgO capsule, a secondary process inferred in Bridgestock et al. (2014). A series of three Fe-Sn + MORB experiments was performed over a temperature range of 1673-1973K to investigate the influence (if any) of Sn on Zn metal-silicate isotope fractionation, as well as to extend our investigation on the effects S to lower temperature by comparison to Sn-bearing experiments. Again, run durations were varied as a function of temperature, with shortened durations at elevated temperatures. A series of three experiments was conducted with the same Fe-FeS starting composition for the metallic phase as the first series, but with a haplogranitic silicate composition (Fe-FeS + HPLG), to extend the investigation of any potential temperature effects to lower temperatures than can be achieved with a basaltic silicate component. Temperatures for the experiments ranged from 1473 to 1673K. Experiments at 1473K and 1573K were first heated to 1773K for 30 minutes, then brought down to the intended temperature of equilibration for an additional 30 minutes. This procedure ensured complete differentiation of the metal and silicate phases (at higher temperature) as well as chemical and isotopic (re)equilibration at lower temperatures. Two additional experiments were conducted using only a MORB phase, one with and one without 1wt% Zn added, at 1673K (30 minute run duration) to assess preferential loss of isotopically light Zn during experimental runs, as well as to provide a baseline composition for silicate phases through EPMA analysis. A final set of two experiments were conducted, with the same FeS + MORB starting composition as the first series, in BN capsules to further assess (and confirm) chemical and isotopic equilibrium and loss of isotopically light Zn.

All experiments were quenched (under pressure) by cutting power to the resistive furnace; sample cell and confining pressure were then slowly relieved to minimize decompression cracking of experimental charges. Metal and silicate phases of experiments were manually separated using optical microscopy and magnets. Metal-silicate segregation was very clean, wherein reliable manual separation of a main metallic spherule from a surrounding homogeneous silicate glass matrix was possible, and the enveloping silicate phase had few stranded metallic globules which could be removed, providing metal and silicate phases devoid of cross-contamination for isotopic analyses.

## 2.3. Chemical and Isotopic Equilibrium

A time series of experiments was chosen to assess both chemical and isotopic metal-silicate equilibration time, wherein a lack of correlation between the Zn metal-silicate elemental and isotope fractionation with time effectively illustrates equilibration, and aberrations from case-normative values at short run durations indicate disequilibrium. This method has been validated for both partitioning (Corgne et al., 2008; Siebert et al., 2011) and isotopic (Bridgestock et al., 2014) equilibrium, and therefore was the preferred method for this study. Chemical equilibrium was further evidenced through observation of metal-silicate segregation upon run completion.

## 2.4. Dissolution and Chemical Purification

Metallic cores and surrounding silicate matrices from experiments were carefully separated mechanically to minimize cross-contamination, then crushed to facilitate dissolution by acid attack. Reliable isolation of micro-sized metallic globules stranded in the silicate matrix was not possible. Metallic samples were dissolved in 6N HCl; silicate samples were sequentially dissolved first in a concentrated $HNO_3$/HF mixture to ensure full digestion of silicate bonds, dried, and the residuum dissolved in 6N HCl. Isolation of Zn ions from other species was achieved through ion exchange chromatography using a strong anion exchange resin (Bio-Rad™ AG1 X8, 200-400 mesh) following a technique outlined elsewhere (Moynier et al., 2006; Moynier and Le Borgne, 2015). Resin was washed prior to ion exchange through repeated $HNO_3$ and Milli-Q water rinsing. Once loaded in exchange columns, Zn remains adsorbed to the resin as a bromine complex while most other cationic species are eluted with HBr and discarded; Zn is subsequently eluted using dilute $HNO_3$.

## 2.5 Mass Spectrometry (MC-ICPMS)

All experimental samples were dissolved in 0.1N $HNO_3$ for isotopic analyses. Zinc isotope mass spectrometric analyses were performed on these purified solutions using a Thermo Scientific Neptune Plus MC-ICPMS at the Institut de Physique du globe de Paris, outfitted with an SSI quartz nebulizer / spray chamber (see Moynier and Le Borgne, 2015).

Zinc ion intensities were measured using Faraday cups in conjunction with $10^{11}$ ohm resistors. Corrections for instrumental mass bias were performed using sample-standard bracketing with the JMC Lyon Zn reference solution (Marechal et al., 1999); $\delta^{66}Zn$ for all samples are reported relative to JMC Lyon and calculated via Equation 1,

Equation (1) $$\delta^n Zn = \left[ \frac{^nZn/^{64}Zn_{Sample}}{^nZn/^{64}Zn_{JMC\ Lyon}} - 1 \right] * 1000$$

where $n$ = 66 or 68. As all Zn fractionation herein was mass-dependent, only $\delta^{66}Zn$ will be discussed. Zinc isotopic compositions ($\delta^{66}Zn$) of metal and silicate phases from all experimental runs were analyzed multiple times (3-5 replicates for most samples). The average $\delta^{66}Zn$ value and the metal-silicate Zn isotope fractionation, $\Delta^{66}Zn_{met-sil}$ (Equation 2),

Equation (2)            $\Delta^{66}Zn_{met-sil} = \delta^{66}Zn_{metal} - \delta^{66}Zn_{silicate}$

were calculated for each experimental pair. Error for all $\delta^{66}Zn$ values reported are 2σ, and error for all $\Delta^{66}Zn_{met-sil}$ reported are 1σ, in line with convention. External reproducibility and accuracy of the complete experimental procedure was evaluated by processing BHVO-2 and Zn dopant replicate samples through the entire methodology (Table 3). BHVO-2 #1 and #2 represent two different dissolutions of the same homogeneous powder; BHVO-2 #1 was then split in three aliquots that were individually processed through the chemistry. Two times the standard deviation (2σ) for these four replicates of BHVO-2 was 0.02‰ (Table 3), in very good agreement with a previous assessment of external reproducibility by Chen et al. (2013a) of ~0.05‰ or better (2σ). Additionally, $\delta^{66}Zn$ obtained for BHVO-2 (0.30 ±0.02‰) and AGV-2 (0.25 ±0.08‰) are in good agreement with reference values of 0.28 ±0.04‰ and 0.29 ±0.03‰ (2σ) (Moynier et al., 2017), confirming good accuracy for this method (see Tables 2 and 3). Five full procedural replicates from five separate dissolutions containing the Zn dopant used in our experiments were analyzed to further investigate reproducibility when applied to dissolutions of varying composition containing the same Zn source. Zn dopants #3, #4, and #5 were analyzed 3 months after Zn dopants #1 and #2 to account for any annual effects. The 2σ error for these five complete replicates was 0.04‰, further indicating a high level of analytical precision for the entire method. Isotopic data for all experiments are detailed in Tables 2 and 3, and Appendix 1.

## 2.6 Electron Microprobe Analysis (EPMA)

Representative pieces of metal and silicate phases from all experimental runs were mounted and polished in a hardened resin, then carbon coated prior to analysis. All selected pieces were greater than 30 microns in size to ensure reliable analyses. Element concentrations for both phases were analyzed using a CAMECA SX Five electron microprobe at the CAMPARIS facility in Paris (Université Pierre et Marie Curie, Paris). Operating conditions were 15kV accelerating voltage with a 40nA beam current, with counting times of 10-20s on background and peak for major elements and 20-40s on background and peak for minor elements. Minerals and pure oxides were used as standards for all elements, and ZnS was used as a standard for Zn and S. For both metal and silicate phases, averages from multiple (>10) 20-30μm$^2$ beam rasters were used to determine bulk compositions, with all error for EPMA data reported as 2SE. Metallic phases often display heterogeneity along the interface with the surrounding silicate, a result of both nucleation and mass diffusion processes upon quenching (O'Neill et al., 1998 and therefore measurements were confined to the inner ~⅔ of the metallic spherules. Many experiments contained small (<20μm diameter) metallic globules trapped at or near the silicate/capsule interface that failed to coalesce with the main metallic spherule, either as a result of adhesive forces imposed by the capsule or from entrapment in olivine growth caused by MgO enrichment in the silicate melt immediately adjacent to the capsule wall. Due to their small size, homogenization through the integration of raster screens was not possible for stranded metallic globules and it was not possible to sample areas unaffected by boundary heterogeneities, therefore reliable analysis of stranded metal droplets along the silicate/capsule interface was not possible.

## 3. Results

### 3.1 Textures and elemental compositions

All experiments were conducted at conditions exceeding the liquidus of both metal and silicate phases (Fig. 1). Silicate phases for all experiments were glassy in texture, except for runs at high temperatures (above 1700K) where MgO dissolution (from the capsule) into the silicate melt caused olivine saturation along the capsule wall and growth of skeletal olivine upon quenching (Fig. 2a). Metallic phases for all experiments were cryptocrystalline, except for S- and Sn-bearing experiments, both of which displayed quench textures comprised of FeS-rich matrices with interstitial Fe-rich globules that exsolved upon quenching (Fig. 2b), and the like for Sn. The textures of both phases confirm the superliquidus conditions of the experiments. Separation and coalescence of metallic liquids into spherules approximately 1mm in diameter was achieved in all experiments (Fig. 1), with the exception of a few small metal droplets (1-20μm diameter) stranded along the sample-capsule interface (more prominent in experiments that experienced MgO enrichment). Backscattered electron (BSE) images indicated that no "nano-nuggets" of metal were present in the quenched silicate melt, further indicating that there was no cross-contamination between metal and silicate phases for analyses (partitioning and isotopic fractionation).

The major element composition of the basaltic silicate was: $SiO_2$ (47.0 ±0.31wt%), $Al_2O_3$ (14.58 ±0.08wt%), MgO (10.3 ±0.43wt%), CaO (9.29 ±0.05wt%), FeO (7.74 ±0.13wt%), $Na_2O$ (3.18

±0.09wt%). These values are from EPMA measurements of the MORB used for all experiments ('*179*' in Appendix 2), post-experimentation at 1673K for 30 minutes, as at these conditions there is no appreciable enrichment in MgO from the capsule into the silicate phase, and therefore this provides the best baseline basaltic silicate composition for the experimental dataset. The MgO content of the silicate phase ranged from 10wt% to 40wt% as a function of experimental temperature, as higher temperature experiments displayed higher levels of MgO enrichment (see Appendix 2). Major element composition of the metallic phase was predominantly Fe with S contents varying from 0 to 30wt%, and Sn contents varying from 0 to 16wt% (Table 1). Zinc content in both the silicate and metal phase was 0 to 1wt%. Metal and silicate compositions for each experiment (Appendix 2) vary according to compositional suite and experimental conditions. At the scale of the EPMA analyses, both silicate phases and nominally pure Fe metallic phases were compositionally homogeneous. S- and Sn-bearing metallic spherules displayed quench textures. These heterogeneities are smaller than the raster screens (see Fig. 2) and lead to minor variations in composition analyses (i.e. high standard deviation), and were therefore integrated through the averaging of multiple raster screens to give a homogenized bulk composition for the metallic phase (after Chabot et al., 2009; Siebert et al., 2011). Oxygen fugacity relative to the iron-wüstite (IW) buffer was calculated via:

Equation (3) $$\Delta IW = 2\log \frac{(X_{FeO}^{silicate})}{(X_{Fe}^{metal})}$$

where ideal mixing behavior for Fe in both silicate and metallic melts has been assumed, and therefore reported ΔIW values reported here are minima (see Siebert et al., 2011). Within the dataset, ΔIW ranged from -2.68 to -1.55 (Table 4).

**3.2. Isotopic composition of the starting materials**

The $\delta^{66}$Zn values of the MORB, Zn powder, and starting materials for Fe-FeS experiments are reported in Table 2 and Appendix 1. The MORB, Zn powder, and Fe-FeS + MORB starting material displayed values of $\delta^{66}$Zn= +0.34 ±0.03‰, +0.13 ±0.01‰, and +0.15 ±0.05‰, respectively. All starting materials were mixed with 60wt% MORB and 1wt% Zn powder. Zinc abundance in the starting material is dominated by the 1wt% Zn due to the low natural abundance of Zn in MORB (~100ppm; e.g. Herzog et al, 2009), and therefore the starting materials consistently share the same initial $\delta^{66}$Zn signature as the Fe-FeS starting material (0.14 ±0.04‰, 2σ) within error.

### 3.3. Time Series Experiments

A series of Fe-FeS experiments at 1673K and 2 GPa were conducted across a broad span of experimental durations (5-240 minutes). For all time series experiments, $\Delta^{66}$Zn$_{met-sil}$ remained statistically constant and did not vary as a function of time, with values of +0.04 ±0.06‰ to +0.12 ±0.05‰ (Table 2, Fig. 3).

### 3.4. Results for Temperature Experiments

Four compositionally distinct series of experiments were conducted (details in **Section 2.2** and Table 1) to investigate the effects of temperature and composition on the isotopic fractionation of Zn. For all temperature experiments, $\Delta^{66}$Zn$_{met-sil}$ remained statistically constant and did not vary as a function of temperature. Fe-FeS experiments in MgO capsules (including the time series experiments), with temperatures from 1473K to 1973K, displayed $\Delta^{66}$Zn$_{met-sil}$ values of -0.01 ±0.01‰ to +0.12 ±0.05‰ (Table 2, Fig. 4). Fe + MORB experiments, with temperatures from 1973K to 2273K, displayed $\Delta^{66}$Zn$_{met-sil}$ values of 0.00 ±0.05‰ to +0.11 ±0.06‰. Fe-Sn + MORB experiments, with temperatures from 1673K to 1973K, displayed $\Delta^{66}$Zn$_{met-sil}$ values of +0.01 ±0.05‰ to +0.09 ±0.06‰ (Fig. 4). No resolvable dependency of $\Delta^{66}$Zn$_{met-sil}$ on temperature was observed for any experimental series (Fig. 4).

### 3.5. Results for Zn loss experiments

Both metal and silicate phases of all experiments in MgO were isotopically heavier than the measured bulk $\delta^{66}$Zn of the starting materials (~0.14‰) (Tables 2 and 3, Appendix 1), and thus necessarily bulk $\delta^{66}$Zn values of experimental run products (i.e. metal plus silicate) yielded values that were heavier than that of the starting material. Therefore, two MORB-only experiments (one with, one without Zn dopant), at 1673K in MgO capsules (30 minute run duration) were conducted, and also displayed heavier $\delta^{66}$Zn values of +0.20 ±0.01‰ and 0.42 ±0.02‰ (respectively) relative to the starting material (experiments *177* and *179* in Table 2, Appendix 1).

## 3.6. Results for BN experiments

Two additional Fe-FeS +MORB experiments were conducted in BN capsules at 1673K, with run durations of 30 and 60 minutes, to investigate the effects of using a different capsule material, notably any compositional effect of boron oxide ($B_2O_3$) in the silicate phase, and to further investigate chemical and isotopic equilibrium and the preferential loss of isotopically light Zn. $\delta^{66}Zn$ values for both metal and silicate phases for these experiments displayed no preferential loss of isotopically light Zn during experimentation, as they were statistically identical to that of the starting material (Table 2). $\Delta^{66}Zn_{met-sil}$ for these experiments remained constant and did not vary, with values of -0.05 ±0.04‰ and -0.04 ±0.05‰ (30 and 60 minutes, respectively) (Table 2, Fig. 4).

## 3.7. Zn metal-silicate partitioning

### 3.7.1. Formalization of Zn metal-silicate partitioning

The partitioning of Zn between metal and silicate is described by the reaction,

Equation (4) $$Zn_{n/2}^{silicate} + \frac{n}{2} Fe^{metal} = \frac{n}{2} FeO^{silicate} + Zn^{metal}$$

where *n* is the valence state of the Zn cation in the silicate melt (2+) and metal silicate exchange of Zn is related to the reduction-oxidation of Fe.

When equilibrium has been established, the logarithm of the equilibrium constant *K* can be written as:

Equation (5) $$\log K = \log \frac{(X_{FeO}^{silicate})^{n/2} * (X_{Zn}^{metal})}{(X_{ZnO_{n/2}}^{silicate}) * (X_{Fe}^{metal})^{n/2}} + \log \frac{(\gamma_{Zn}^{metal})}{(\gamma_{Fe}^{metal})^{n/2}} + \log \frac{(\gamma_{FeO}^{silicate})^{n/2}}{(\gamma_{ZnO_{n/2}}^{silicate})}$$

where the first term on the right side of the equation, $K_M^D$ or the exchange coefficient, can be used as a means to elucidate the effects of temperature, pressure, and where the effect of oxygen fugacity has been considered by normalizing the data to the partitioning of Fe. Following Wade and Wood (2005), oxide activity coefficient ratios were considered constant (as Zn is a low valence cation), and thus the equilibrium constant used to express partitioning results can be written as:

Equation (6) $$\log K_e = \log K_M^D + \log \frac{(\gamma_{Zn}^{metal})}{(\gamma_{Fe}^{metal})^{n/2}}$$

where $K_e$ is the effective equilibrium constant, and $K_M^D$ is experimentally derived and activity coefficients are calculated as described below. The presence of both C and S in the metal phase is known to influence the partitioning of trace elements (e.g. Chabot and Agee, 2003 for carbon); when applicable, corrections were made for the influence on Zn partitioning following Siebert et al. (2011) for C and Wood et al. (2014) for S, where activity coefficients were determined using the interaction parameter approach and method described by Ma (2001). This approach allows the use of tabulated interaction parameters $\varepsilon$ (The Japan Society for the Promotion of Science and the Nineteenth Committee on Steelmaking, 1988) and quantitatively expresses the thermodynamics of metallic solutions. In this study, $\varepsilon_{Zn}^C = 5.7$ was used for the interaction of C (updated from Siebert et al., 2011) and $\varepsilon_{Zn}^S = -1.7$ was used for the interaction of S, as given by Wood et al. (2014). The influence of silicate melt composition has been considered negligible within the dataset as the low valence state of Zn renders it largely insensitive to the structure and composition of the silicate melt (Siebert et al., 2011).

### 3.7.2. Exchange coefficients for Zn as a function of reciprocal temperature

Exchange coefficients (Table 4) displayed an increasing trend with increasing temperature (Fig. 5a), in very good agreement with previous work (e.g. Siebert et al., 2011). In order to develop the most robust regression model for Zn partitioning to date, complimentary datasets were input with the same corrections for C and S (where applicable) as our own dataset, and have been reported in Fig. 5b along with our multilinear regression model (calculated as described in *Section 4.3.1*). To fully parameterize the partitioning of Zn as function of temperature and pressure, data across broad ranges of these variables have been included from this study, Lagos et al. (2008), Siebert et al. (2011), Ballhaus et al. (2013), Wood et al. (2014), and Wang et al. (2016) (Appendix 3). In general, all Zn data from the literature are in good agreement and have been included in our regression model. However it should be noted that data from Mann et al. (2009) displayed notable discrepancies (especially at high pressure) with the rest of the data in regards to the effect of pressure on Zn partitioning, and for that reason, data from the study has been excluded due to an inability to reconcile these differences.

## 4. Discussion

### 4.1. Chemical/isotopic equilibrium and Zn loss

Achievement of compositional and isotopic equilibrium between metal and silicate phases is requisite for an accurate characterization of elemental partitioning and isotopic fractionation. The Fe-FeS time series experiments presented herein displayed no systematic correlation between $\Delta^{66}Zn_{met-sil}$ and run duration, as the magnitude of Zn metal-silicate isotope fractionation at 5 min and 240 minutes was effectively equal (Fig. 3). The constancy of $logK_e$ regardless of run duration further indicates that Zn compositional and isotopic equilibrium was attained early in the experimental runs (Table 3, Fig. 5a). This indicates that both compositional and isotopic equilibrium were established very early (≤10 minutes) at temperatures at and above 1673K, further constraining the previously estimated equilibration time of approximately one hour (Bridgestock et al., 2014). More importantly, $\Delta^{66}Zn_{met-sil}$ for the experiments did not vary with time, implying that loss of isotopically light Zn did not lead to *apparent* Zn metal-silicate isotopic fractionation induced by preferential diffusional loss of isotopically light Zn into MgO capsules, and therefore did not affect $\Delta^{66}Zn_{met-sil}$. Additionally, Zn was undetectable in the capsule material via EPMA (~200ppm detection limit) and therefore the amount of Zn diffused into the capsule was considered negligible with respect to the Zn reservoir in the experiments (~8mg).

For all experiments in MgO capsules, as the experiments of Bridgestock et al. (2014) in MgO capsules, the $\delta^{66}Zn$ of both metal and silicate phases, and necessarily bulk $\delta^{66}Zn$ (mass balance calculations using Table 1) were isotopically heavier than the starting material, indicating preferential loss of isotopically light Zn via diffusional mass transfer. Complimentary experiments in MgO were run in order to directly address this loss from experimental charges during experimentation. Our Zn loss experiments, which provide comparison between the isotopic signatures of bulk materials before and after experiments (see Table 2), confirmed the presence of this secondary process, and comparison of these results to our time series experiments provided a means to assess preferential loss of isotopically light Zn as a function of time. Although $\delta^{66}Zn$ values appear to qualitatively increase as a function of run duration (Table 2), no statistical trend was found between bulk $\delta^{66}$Zn (individual or bulk), and therefore $\Delta^{66}$Zn$_{met-sil}$, and time. Moreover, and perhaps the most convincing validation of chemical and isotopic equilibrium, our experiments in BN capsules - the results of which are statistically identical to those in MgO capsules in regards to partitioning and isotopic fractionation in our study - displayed no measurable preferential loss of isotopically light Zn (i.e. closely approximated a closed system) and no correlation between $\Delta^{66}$Zn$_{met-sil}$ and time, further indicating that chemical and isotopic equilibrium was achieved within the experimental dataset.

### 4.2. Zn metal-silicate isotopic fractionation
#### *4.2.1. Systematics of $\Delta^{66}$Zn$_{met-sil}$*

The Zn metal-silicate isotope fractionation factor, $\Delta^{66}$Zn$_{met-sil}$, regardless of experimental conditions (T, X, fO$_2$), was between -0.05 ±0.01‰ and +0.12 ±0.04‰ within our study, indicating that the sensitivity of Zn isotopic fractionation to the experimental conditions of this study was negligible. No statistically robust trend was found between $\Delta^{66}$Zn$_{met-sil}$ and temperature, indicating that the sensitivity of Zn isotopic fractionation to temperature is not resolvable at viable conditions of metal-silicate differentiation between molten phases. Our dataset is in general accord with the $\Delta^{66}$Zn$_{met-sil}$ values of Bridgestock et al. (2014), which ranged from -0.14 ±0.12‰ to +0.21 ±0.15‰; our data reaffirm the absence of Zn metal-silicate isotope fractionation in both datasets within analytical uncertainty, and further constrain both equilibration time and analytical precision (see Fig. 3). The agreement in $\Delta^{66}$Zn$_{met-sil}$ between S-bearing, Sn-bearing, and nominally S- and Sn-free metallic phases, suggests that both elements have a negligible effect on Zn metal-silicate isotope fractionation. The constancy of $\Delta^{66}$Zn$_{met-sil}$ regardless of silicate composition further suggests that Zn metal-silicate isotopic fractionation is largely independent of silicate melt composition. Furthermore, $\Delta^{66}$Zn$_{met-sil}$ displayed no dependence on redox state within the fO$_2$ range of this study (ΔIW values from -2.68 to -1.55).

In detail, the observations of the current study suggest that $\Delta^{66}$Zn$_{met-sil}$ at temperatures at or near that of core formation of the Earth, estimated to be in excess of 3500K (Siebert et al., 2013), would be effectively zero. Even at the low temperatures (1473K) achieved in this study no resolvable fractionation was observed. Pressure is generally theorized to have a negligible effect on isotopic fractionation via changes in molar volume (Criss, 1999; Schauble, 2004), however changes in the stiffness of force constants in contracting media can have observable isotopic effects at high pressure (Shahar et al., 2016), but this requires in-situ synchrotron measurements on solids. In regards to Zn metal-silicate isotopic fractionation, a comparison of previous studies at lower pressure (Moynier et al., 2005; Bridgestock et al., 2014) and our data at 2 GPa has thus far shown no resolvable effect of pressure on $\Delta^{66}Zn_{met-sil}$ at or below 2 GPa. Redox conditions have been shown to affect Zn partitioning (Corgne et al., 2008; Siebert et al., 2011), and can play an important role in isotopic fractionation through changes in the oxidation state of the element of interest (e.g. Schauble, 2004) and/or through changes in the bonding environment of the silicate melt phase (Shahar et al., 2008; Dauphas et al., 2014). However, previous studies (e.g. Siebert et al., 2011) have determined that Zn stays in the 2+ oxidation state over the range of possible $fO_2$ conditions during core formation in Earth (-3.5<$\Delta$IW<-1.5). Moreover, no resolvable effect of redox state on $\Delta^{66}Zn_{met-sil}$ was observed over the similar range of $fO_2$ conditions of our experiments (-2.68<$\Delta$IW<-1.55), suggesting that $\Delta^{66}Zn_{met-sil}$ is insensitive to $fO_2$ at conditions relative to Earth, or even most smaller planetary bodies (e.g. Wadhwa, 2008). All of these factors, together with the negligible effect of temperature observed in our study, implies that core formation processes, even in smaller planetary bodies (i.e. lower T and P, varying $fO_2$), is unlikely to produce resolvable Zn metal-silicate isotopic fractionation.

### 4.2.2. $\Delta^{66}Zn_{met-sil}$ and implications for planetary bodies

The dataset suggests that for most differentiated bodies, e.g. planetesimals and larger, the $\delta^{66}Zn$ of silicate phases is a viable proxy for the bulk $\delta^{66}Zn$ of the differentiated body, as disparities between the two values (from metal-silicate differentiation) would be within analytical uncertainty. Chen et al. (2013b) have suggested a value of $\delta^{66}Zn$ = +0.28 ±0.05‰ for BSE ($\delta^{66}Zn_{BSE}$, based on measurements of basalts) and furthermore suggested that the $\delta^{66}Zn$ of the bulk Earth ($\delta^{66}Zn_{BE}$) would be indistinguishable from this value by assuming no partitioning of Zn into the core. Although the current study indicates that a large fraction of Zn present at the time of core/mantle differentiation would partition into the core, the use of $\delta^{66}Zn_{BSE}$ (basaltic or not) for $\delta^{66}Zn_{BE}$ remains valid, as at the high temperatures of core formation Zn metal-silicate isotope fractionation would be essentially nil. The same logic follows for other large differentiated bodies such as the Moon, validating the use of $\delta^{66}Zn$ of lunar basalt as a proxy for $\delta^{66}Zn$ of the bulk Moon. In addition, the comparatively minor effects of Zn isotope fractionation during both core formation and igneous processes (both smaller than ~0.10‰, present study and Chen et al. (2013b), respectively) would be overpowered by Zn isotope fractionation during volatilization, *c.f.* the +1‰ deviation of lunar basalts from terrestrial basalts observed by Paniello et al. (2012) and Kato et al. (2015).

**4.3. Zn metal-silicate partitioning**

*4.3.1. Thermodynamics of Zn metal-silicate partitioning*

In order to isolate and describe the various thermodynamic controls on partitioning, it is useful to express the partitioning behavior of Zn through the relation,

Equation (7) $$\log K_e = a + \frac{b}{T} + \frac{c*P}{T} + d*nbo/t$$

where $a$, $b$, $c$, and $d$ are regression constants determined via least squares multivariable regression, $T$ is temperature in K, $P$ is pressure in GPa, and $nbo/t$ is the molar ratio of non-bridging oxygens over tetrahedral cations in the silicate melt. This method follows previous studies (Righter et al., 1997; Wade and Wood, 2005; Corgne et al., 2008; Siebert et al., 2011) and further details can be found therein. Assuming that the partitioning of Zn is effectively independent of silicate composition, this expression can be simplified to,

Equation (8) $$\log K_e = a + \frac{b}{T} + \frac{c*P}{T}$$

Using this expression for the logarithm of the equilibrium constant, the regression constant $a$ is the intercept, and constants $b$ & $c$ effectively describe the sensitivity of partitioning to temperature and pressure (respectively). Partitioning data (and T, P) from this study (Table 4) have been combined with data from literature (Lagos et al., 2008; Siebert et al., 2011; Ballhaus et al., 2013; Wood et al., 2014; Wang et al., 2016) (Appendix 3) in order to provide the most robust regressional analysis of Zn metal-silicate partitioning to date ($N = 78$). Our regression of this data compilation yields regression constants of $a = 0.69 \pm 0.12$ (dimensionless), $b = -3500 \pm 228$ ($K$), and $c = -16 \pm 10$ ($K/GPa$) (error reported as 2σ uncertainties). These results illustrate that metal-silicate partitioning in Zn is sensitive to temperature, while the pressure effect remains statistically unresolvable.

*4.3.2. Modeling core formation/differentiation*

One of the most relevant and useful applications of the equilibrium constant as a function of the thermodynamic variables that affect it is the use of Equation 8 (with regression constants) to evaluate the validity of various core formation models. Here we have evaluated various continuous stage core formation models, as single stage models are physically implausible for Earth and often fail to recreate current mantle element abundances (e.g. Ga, Ge, P) under viable P-T-redox conditions, as solutions often require P-T conditions that are below (or above) the peridotite solidus (e.g. Wade and Wood, 2005; Siebert et al., 2011).

To that end, we have evaluated three continuous core formation models in which both temperature and pressure increase as a function of the accretion process, and oxygen fugacity either remains constant, decreases (oxidizing, denoting initial accretion conditions), or increases (reducing) with the fraction accreted. Accretion is modeled in 1% increments (Wade and Wood, 2005), temperatures were bound by the peridotite liquidus (averaged after Andrault et al. 2011 and Fiquet et al. 2010), and pressure of equilibration in the magma ocean was set as 40% of the core-mantle boundary value for each accretion increment (Siebert et al., 2012). Considering constant oxygen fugacity during continuous core formation, or accretion under oxidizing conditions and progressive reduction of the primitive mantle by oxygen incorporation to the Earth's core (Siebert et al., 2013), leads respectively to partition coefficients for Zn ($D^{Zn}_{Core-Mantle}$) between 2.7 ±1.2 and 1.7 ±0.7 (error for $D^{Zn}_{Core-Mantle}$ is propagated through the model as 2σ uncertainties from regression constants). Following numerous previous works (e.g. Wade and Wood, 2005; Corgne et al., 2008), in the case of the reducing model oxygen fugacity of the silicate mantle progressively increases during accretion from reducing conditions (ΔIW-3.3) to the current core-mantle equilibrium value (ΔIW-2.3; i.e. 8wt% FeO in the mantle) through heterogeneous accretion (Rubie et al., 2011) or self-oxidation processes involving Si, Fe, and O with a single homogeneous accreting composition (Wade and Wood, 2005; Corgne et al., 2008; Javoy et al., 2010). The resultant partitioning coefficient of Zn, $D^{Zn}_{Core-Mantle}$, for this model is 4.5 ±2.0 (2σ). Taking into account all cases and associated errors yields a range of values for Zn partitioning between 1.0 and 6.5. Therefore, unless one considers only the lowest possible value for $D^{Zn}_{Core-Mantle}$, in the case of oxidizing conditions with maximum negative error, the Earth's core most likely holds a large fraction of its Zn, and may possibly be its dominant Zn reservoir.

### 4.3.3. The Missing Zn – Harmonizing $D_{met-sil}^{Zn}$ and iron meteorites

It is necessary to compare our experimental results to a natural system in order to assess our parameterization of $D_{met-sil}^{Zn}$. Iron meteorites offer perhaps the most accessible natural analogue for Zn metal-silicate partitioning, as the magmatic types are interpreted to be the cores of differentiated parent bodies (e.g. Goldstein et al., 2009). Interestingly, Zn contents in iron meteorites vary between 0.1ppm and 45ppm (e.g. Luck et al., 2005; Chen et al. 2013a) an observation which is seemingly at odds with the experimentally derived partition coefficient for Zn in Bridgestock et al. (2014).

Using our parameterization of Zn metal-silicate partitioning and estimated conditions of core formation for iron meteorite parent bodies (P, T, fO$_2$), we can predict a range of Zn concentrations for iron meteorites based on plausible conditions of parent body formation/differentiation. The pressure of core formation in iron meteorite parent bodies is theorized to range from essentially atmospheric (Righter and Drake, 1996) to less than 0.2 GPa for the largest asteroidal parent bodies (McCoy et al., 2006). Since pressure is observed to have a negligible effect on $D_{met-sil}^{Zn}$, and since both the Fe-FeS eutectic and the melting point of pure Fe do not change considerably between 0-3GPa, the temperature of core formation can effectively be bracketed between 1271K and ~1900K (Buono and Walker, 2011), with a higher temperature being more likely as it allows for sufficient melting for core-mantle segregation and equilibration (Righter and Drake, 1996). The redox conditions of core formation in iron meteorite parent bodies are not well known, therefore a range of reasonable values (-3<ΔIW<-1) that satisfy many current observations and oxygen fugacities for hypothesized parent bodies in the literature can be used (e.g. Righter and Drake, 1996; Burbine et al., 2002; Wadhwa, 2008; Ermakov et al., 2014; Neeley et al., 2014; Consolmagno et al., 2015). In accordance with these studies with respect to parent body composition, Zn contents of 1ppm, 50ppm, and 100ppm were chosen to encompass representative Zn concentration values for various parent body compositions (volatile depleted, H chondrite, or Enstatite chondrite type parent body). S contents in iron meteorites are not well-constrained as they cannot be inferred from bulk measurements (Goldstein et al., 2009); for this reason and for model simplicity, the influence of S has not been included (i.e. zero S in the parent body); however, it is noted that S will increase the partitioning of Zn into the metallic phase (e.g. Wood et al., 2014). Assuming a mass fraction for the parent body core of 0.3 will provide an upper bound (via mass balance) for Zn abundance estimates in iron meteorites (and qualitatively compensate for the zero S solutions). Using Equation 6 and our parametric solution to Equation 8, $D^{Zn}_{Core-Mantle}$ for the parent body is calculated as a function P,T, and fO$_2$, and used in conjunction with Zn concentrations in the three representative parent bodies to estimate the Zn contents expected in the differentiated core of the parent body, i.e. iron meteorites. Using this approach, the estimated $D^{Zn}_{Core-Mantle}$ at IW-1 and low temperature (1271K) is ~0.03, yielding lower bound Zn content estimates of ~0.04ppm, 2ppm, and 4ppm in the three respective parent body cores. Under more reducing conditions (IW-3) and at higher temperatures (1900K), where more Zn is driven into the metallic phase, the estimated $D^{Zn}_{Core-Mantle}$ is ~2, yielding upper limit Zn contents of 2ppm, 81ppm, and 163ppm. Excluding the latter and most extreme case, these results are in very good agreement with the variations in Zn contents of iron meteorites observed, which span across nearly the same range (e.g. Chen et al., 2013a). Thus, the low Zn concentrations observed in iron meteorites can be partly explained by the sensitivity of $D^{Zn}_{met-sil}$ to temperature and fO$_2$ alone.

Furthermore, Chabot et al. (2009) posited that Cr abundance estimates in iron meteorites are likely too low due both to sampling bias, an argument similar to one offered to explain lower-than-expected Si contents in iron meteorites by Pack et al. (2011), and to separation/extraction of chromite from the liquid metallic phase during cooling and crystallization due to the positive buoyancy of chromite crystals in the melt (see reference for further detail), which may further decrease measured Zn contents in iron meteorites. In line with this notion, Bridgestock et al. (2014) determined that part of the Zn in iron meteorites is sequestered in minor chromite inclusions, as their largest measured concentration of Zn was in a chromite phase from the Toluca iron meteorite, providing tentative observational evidence for such a mechanism. Therefore, if sampling bias and/or Zn sequestration into chromite is taken into consideration, these estimated values will be even lower, especially considering that 0.3<$D^{Zn}_{solid\ metal-liquid\ metal}$<0.9 (Chabot et al., 2009; Rai et al., 2013, respectively), and thus the upper limit of our calculations (averaged enstatite chondrite composition at low fO$_2$ and high T) may fall into accord with natural observations.

Taken as a whole, our results indicate that although sampling bias and sequestration of Zn into minor phases most likely occur to some degree, the low Zn concentrations in iron meteorites can be largely explained by the dependence of $D_{met-sil}^{Zn}$ on temperature and fO$_2$ regardless of the parent body composition. With this in mind, sampling bias and Zn sequestration into minor phases (e.g. chromite) need only be invoked given a parent body Zn concentration of 100ppm or greater and/or highly reducing conditions.

### *4.3.4. Upper estimate for Zn content of bulk Earth*

Regardless of which core formation model is applied (***Section 4.3.2***), our new estimates for the coefficient of distribution for Zn is significantly larger than previous geochemical estimates of Zn core/mantle partitioning (e.g. Dreibus and Palme, 1996), indicating that the core may hold a large fraction of Earth's Zn, especially in the case of continuous stage core formation under reducing conditions, which yields a $D_{met-sil}^{Zn}$ (core-mantle) of 4.5 ±2.0. Using this newly constrained metal-silicate partitioning coefficient for Zn and the Zn content of the Earth's mantle (BSE, at 53.5 ±2.7ppm)  (Palme and O'Neill, 2003), we can then estimate the Zn contents of the Earth's core and that of the bulk Earth. Using this approach, we have re-estimated the Zn content of the Earth's core as 242 ±107ppm, which then yields a bulk Earth Zn content of 114 ±34ppm. With previous calculations between 24ppm (Allègre et al., 2001) and 47ppm (Kargel and Lewis, 1993), this new estimation significantly increases the Zn budget of the bulk Earth, even at the lower bound (65ppm) of our estimation.

### *4.3.5. Upper estimate for S content in the core*

Sulfur is a siderophile and moderately volatile element with a 50% condensation temperature close to that of Zn (T$_{50}$ = 664K), as such its abundance in the bulk Earth is related to the abundances of Zn and similarly volatile elements. Using a similar approach to Dreibus and Palme (1996), we have recalculated the S content in the Earth's core using updated values for the Zn and S abundances in the BSE and CI carbonaceous chondrites (as an analogue for the chemical composition of the bulk Earth). Abundances in the BSE for Zn (53.5 ±2.7ppm) and S (0.02 ±0.01wt%) are from Palme and O'Neill (2003). The CI abundance for Zn (303 ±6ppm) is from Barrat et al. (2012) and the CI abundance for S (5.410 ±0.365wt%) is from Lodders (2003). Dreibus and Palme (1996) reported a maximum S content in the Earth's core of 1.7wt%, which has been accepted as the nominal value for nearly two decades. A recent estimate from geochemical constraints (Savage et al., 2015) yielded a lower S content for the core to be a minimum of 0.6wt%, and estimates from sound speed velocity and density studies (e.g. Badro et al., 2007; Morard et al., 2013) range from 0wt% to ~6wt% S in the core (respectively).

The S content of the Earth's core that Dreibus and Palme (1996) calculated is based on the assumption that the Zn content of the mantle is equal to that of the bulk Earth, i.e. that Zn is completely lithophile and therefore none resides in the core. The findings of our study suggest that this is not the case and that in fact the Earth's core likely hosts a higher concentration of Zn than the BSE, and thus the bulk Earth holds appreciably more Zn than the mantle alone. Therefore, the estimate of the S content in the Earth's core by Dreibus and Palme (1996) can be revisited by using the updated Zn content of the bulk Earth (114 ±34ppm) from this study in conjunction with the BSE abundances for Zn and S reported above from Palme and O'Neill (2003), a chondritic S/Zn ratio of 179 (based on CI abundances of S and Zn from Lodders (2003) and Barrat et al. (2012), respectively), and a mass fraction of the Earth's core of 0.32 (Siebert et al., 2011). These parameters yield an S content for the bulk Earth of ~2.0 ±0.6wt%, which corresponds to an upper bound S content for the Earth's core of 6.3 ±1.9wt% (in the case of reducing condition). Using different types of carbonaceous chondrites (179<S/Zn<245) or even enstatite (EH) chondrites (S/Zn~200) (Lodders and Fegley, 1998) would not change this estimate significantly and would give a range for the S content of the core up to 8.6wt%. On the other hand, ordinary chondrites have significantly higher S/Zn ratio (~400), which would produce unrealistically high amounts of S in the core (>13wt%).

Our geochemical estimate is roughly four fold that of the 1.7wt% S content of the Earth's core as determined by Dreibus and Palme (1996), who did not account for Zn in the core. This estimate is in accord with that of Morard et al. (2013), which estimated ~6wt% S in the core based on high pressure density measurements of liquid iron alloys, but there are some discrepancies with other estimates produced through alternative methodological approaches. For instance, S contents in the core of more than 2wt% largely disagree with current seismological constraints on the geophysical properties of the Earth's core (e.g. Badro et al., 2007; Badro et al., 2014). If the S/Zn ratio for Earth is chondritic, then our results imply that S is a major constituent as a light element in the Earth's core. However, our results can also be interpreted as evidence that the S/Zn ratio for Earth is non-chondritic, which would suggest that either the S/Zn ratio was altered by evaporation/condensation processes prior to (or during) the formation of the Earth, or that Earth's parent body composition itself is non-chondritic (e.g. Wang et al., 2016; Fitoussi et al., 2016).

## 5. Conclusions

We have conducted a suite of experiments to characterize the metal-silicate isotopic fractionation and elemental partitioning of Zn as a function of multiple controls (temperature, pressure, composition, oxygen fugacity). Our study has expanded on the work of Bridgestock et al. (2014), reaffirming that Zn metal-silicate isotopic fractionation is negligible with regard to other processes at temperatures relevant to planetary formation. These results further validate the use of $\delta^{66}Zn_{BSE}$ as a proxy for $\delta^{66}Zn_{BE}$ as suggested by Chen et al. (2013b), even if the bulk of Earth's Zn resides in the core, and furthermore validates this approach for other large differentiated bodies. We have characterized Zn metal-silicate partitioning as a function of temperature, pressure, and composition and determined the range of possible Zn metal-silicate partition coefficients for the Earth to be between 1.6 ±0.7 and 4.5 ±2.0. To investigate the low Zn contents observed in iron meteorites, we applied our parameterization of Zn metal-silicate partitioning for viable conditions of core segregation and differentiation of iron meteorite parent bodies for various representative bulk Zn concentrations to predict their core (iron meteorite) Zn contents, the results of which are in very good agreement with observed iron meteorite Zn abundances. Assuming a continuous stage core formation model for Earth, where the planet becomes increasingly oxidized up to its present state - yielding a partition coefficient of 4.5 ±2.0 - we then re-calculated the Zn budget for the bulk Earth (114 ±34ppm) and the core (242 ±107ppm) assuming reducing conditions of formation (an upper bound). Using these estimates, and a chondritic S/Zn ratio, we then re-calculated an upper limit S content of the Earth's core to be 6-8wt%.

**Acknowledgements**

The authors very kindly thank *GCA* Associate Editor Stefan Weyer and our three anonymous reviewers, whose invaluable input has greatly improved the quality of the manuscript. We thank Paolo Sossi, James Day, and Bruce Fegley for enlightening discussions that have furthered the value of the manuscript. We also thank Nicolas Wehr for his expertise in the petrology lab, as well as Michel Fialin and Nicolas Rividi at the CAMPARIS facility for their expertise with the SX Five microprobe. BM and EP acknowledge financial support from the IDEX SPC for PhD fellowships. JS and BM acknowledge financial support from the French National Research Agency (ANR project VolTerre, grant no. ANR-14-CE33-0017-01). FM and JS thank the financial support of the UnivEarthS Labex program at Sorbonne Paris Cité (ANR-10-LABX-0023 and ANR-11-IDEX-0005-02). FM acknowledges funding from the European Research Council under the H2020 framework program/ERC grant agreement #637503 (Pristine), as well as the ANR through a chaire d'excellence Sorbonne Paris Cité.

constraints from V, Cr, and Mn. *Geochim. Cosmochim. Acta* **67**, 2077–2091.

Chabot N.L., Draper D.S., Agee C.B. (2005) Conditions of core formation in the Earth: Constraints from nickel and cobalt partitioning. *Geochim. Cosmochim. Acta* **69**, 2141–2151.

Chabot N.L, Saslow S.A., McDonough W.F., Jones J.H. (2009) An investigation of the behavior of Cu and Cr during iron meteorite crystallization. *Meteoritics and Planetary Science* **44**, 505-519.

Chambers J.E. (2004) Planetary accretion in the inner solar system. *Earth Planet. Sci. Lett.* **223**, 241–252.

Chen H., Nguyen B.M., Moynier F. (2013a) Zinc isotopic composition of iron meteorites: absence of isotopic anomalies and origin of the volatile element depletion. *Meteoritics and Planetary Science* **48**, 2441-2450.

Chen H., Savage P.S., Teng F.-Z., Helz R.T., Moynier F. (2013b) Zinc isotope fractionation during magmatic differentiation and the isotopic composition of the bulk Earth. *Earth Plan. Sci. Lett.* **369-370**, 34-42.

Consolmagno G.J., Golabek G.J., Turrini D., Jutzi M., Sirono S., Svetsov V., Tsiganis K. (2015) Is Vesta an intact and pristine protoplanet? *Icarus* **254**, 190-201.

Corgne A., Keshav S., Wood B.J., McDonough W.F., Fei Y. (2008) Metal-silicate partitioning and constraints on core composition and oxygen fugacity during Earth accretion. *Geochim. Cosmochim. Acta* **72**, 574–589.

Criss R.E. (1999) *Principles of stable isotope distribution*. Oxford Univ. Press; Oxford.

Dauphas N., Roskosz M., Alp E.E., Neuville D.R., Hu M.Y., Sio C.K., Tissot F.L.H., Zhao J., Tissandier L., Medard E., Cordier C. (2014) Magma redox and structural controls on iron isotope variations in Earth's mantle and crust. *Earth Planet. Sci. Lett.* **398**, 127–140.

Day J.M.D. and Moynier F. (2014) Evaporative fractionation of volatile stable isotopes and their bearing on the origin of the Moon. *Philos. Trans. R. Soc. A* **372**, 20130259.

Dreibus G. and Palme H. (1996) Cosmochemical constraints on the sulfur content in the Earth's core. *Geochim. Cosmochim. Acta* **60**, 1125–1130.

Ermakov A.I., Zuber M.T., Smith D.E., Raymond C.A., Balmino G., Fu R.R., Ivanov B. (2014) Constraints on Vesta's interior structure using gravity and shape models from the Dawn mission. *Icarus* **240**, 146-160.

Fiquet G., Auzende A.L., Siebert J., Corgne A., Bureau H., Ozawa H., Garbarino G. (2010)

**Tables**

Table 1 – Composition of Starting Materials

| Starting Material | Fe-FeS + MORB | Fe + MORB | Fe-Sn + MORB | Fe-FeS + HPLG[a] |
|---|---|---|---|---|
| *Bulk Metal* | 20wt% Fe<br>20wt% FeS | 40wt% Fe | 24wt% Fe<br>16wt% Sn | 20wt% Fe<br>20wt% FeS |
| *Bulk Silicate* | 60wt% MORB | 60wt% MORB | 60wt% MORB | 60wt% HPLG[a] |

[a] HPLG (abbrev. haplogranite) oxides : $SiO_2$ (78wt%), $Al_2O_3$ (11wt%), $K_2O$ (6.4wt%), $Na_2O$ (4.6wt%) (from Tuttle and Bowen, 1958)

Table 2 – Zn isotopic signature ($\delta^{66}Zn$ and $\Delta^{66}Zn_{met-sil}$) of experimental runs by MC-ICPMS

| Starting Material | Experiment | Pressure Media | T (K) | t (min) | $\delta^{66}Zn_{met}$ | 2σ | n | $\delta^{66}Zn_{sil}$ | 2σ | n | $\Delta^{66}Zn_{met-sil}$ | σ[a] |
|---|---|---|---|---|---|---|---|---|---|---|---|---|
| **Fe-FeS + MORB** | *162* | Talc-Pyrex | 1673 | 5 | 0.28 | 0.08 | 4 | 0.24 | 0.08 | 4 | 0.04 | 0.06 |
| | *163* | Talc-Pyrex | 1673 | 15 | 0.28 | 0.04 | 3 | 0.17 | 0.09 | 8 | 0.12 | 0.05 |
| | *158* | Talc-Pyrex | 1673 | 30 | 0.30 | 0.12 | 5 | 0.20 | 0.10 | 6 | 0.10 | 0.08 |
| | *161* | Talc-Pyrex | 1673 | 60 | 0.28 | 0.14 | 5 | 0.25 | 0.08 | 4 | 0.04 | 0.08 |
| | *M2* | Talc-Pyrex | 1673 | 120 | 0.29 | 0.06 | 6 | 0.24 | 0.04 | 6 | 0.05 | 0.04 |
| | *M1[b]* | Talc-Pyrex | 1673 | 180 | 0.43 | 0.05 | 6 | 0.32 | 0.02 | 6 | 0.10 | 0.03 |
| | *M3* | Talc-Pyrex | 1673 | 240 | 0.44 | 0.05 | 6 | 0.38 | 0.07 | 6 | 0.05 | 0.04 |
| | *170* | BaCO$_3$ | 1673 | 120 | 0.65 | 0.07 | 8 | 0.16 | 0.06 | 4 | 0.49[c] | 0.05 |
| | *239[d]* | Talc-Pyrex | 1673 | 30 | 0.13 | 0.04 | 6 | 0.18 | 0.07 | 6 | -0.05 | 0.04 |
| | *240[d]* | Talc-Pyrex | 1673 | 60 | 0.15 | 0.09 | 5 | 0.19 | 0.05 | 5 | -0.04 | 0.05 |
| | *166* | Talc-Pyrex | 1823 | 15 | 0.21 | 0.12 | 8 | 0.21 | 0.01 | 3 | 0.00 | 0.06 |
| | *174* | BaCO$_3$ | 1973 | 3 | 0.25 | 0.05 | 4 | 0.17 | 0.02 | 3 | 0.07 | 0.03 |
| **Fe-FeS + HPLG** | *192[e]* | Talc-Pyrex | 1473 | 30 | 0.33 | 0.03 | 5 | 0.24 | 0.03 | 5 | 0.09 | 0.02 |
| | *189[e]* | Talc-Pyrex | 1573 | 30 | 0.31 | 0.01 | 5 | 0.29 | 0.02 | 5 | 0.02 | 0.01 |
| | *181* | Talc-Pyrex | 1673 | 30 | 0.22 | 0.02 | 5 | 0.23 | 0.02 | 5 | -0.01 | 0.01 |
| **Fe + MORB** | *160* | BaCO$_3$ | 1973 | 5 | 0.21 | 0.08 | 6 | 0.21 | 0.12 | 5 | 0.00 | 0.07 |
| | *167* | BaCO$_3$ | 1973 | 3 | 0.20 | 0.09 | 6 | 0.20 | 0.02 | 3 | 0.00 | 0.05 |
| | *168* | BaCO$_3$ | 2123 | 1 | 0.20 | 0.10 | 6 | 0.12 | 0.07 | 4 | 0.08 | 0.06 |
| | *165* | BaCO$_3$ | 2273 | 1 | 0.25 | 0.09 | 6 | 0.14 | 0.08 | 6 | 0.11 | 0.06 |
| **Fe-Sn + MORB** | *171* | Talc-Pyrex | 1673 | 30 | 0.24 | 0.09 | 4 | 0.22 | 0.07 | 4 | 0.03 | 0.05 |
| | *173* | Talc-Pyrex | 1823 | 15 | 0.27 | 0.09 | 3 | 0.18 | 0.09 | 4 | 0.09 | 0.06 |
| | *172* | BaCO$_3$ | 1973 | 3 | 0.20 | 0.07 | 4 | 0.19 | 0.07 | 4 | 0.01 | 0.05 |
| MORB (undoped) | *177[f]* | Talc-Pyrex | 1673 | 30 | --- | --- | --- | 0.42 | 0.04 | 5 | --- | --- |
| MORB | *179[f]* | Talc-Pyrex | 1673 | 30 | --- | --- | --- | 0.20 | 0.03 | 5 | --- | --- |
| MORB (undoped) | --- | --- | --- | --- | --- | --- | --- | 0.34 | 0.03 | 5 | --- | --- |
| Zn dopant | --- | --- | --- | --- | 0.13 | 0.01 | 5 | --- | --- | --- | --- | --- |
| MORB (mix) | --- | --- | --- | --- | --- | --- | --- | 0.15 | 0.02 | 5 | --- | --- |
| Fe-FeS + MORB | --- | --- | --- | --- | --- | --- | --- | 0.15 | 0.05 | 11 | --- | --- |

[a] Standard deviation (1σ) has been used as the conventional unit reported for $\Delta^{66}Zn_{met-sil}$ (2σ for $\delta^{66}Zn$).

[b] Starting material for experiment *M1* was 60wt% MORB, 8wt% Fe, and 32wt% FeS (bulk doped with 1wt% Zn).

[c] Experiment *170* displayed an abnormally large $\Delta^{66}Zn_{met-sil}$ value, interpreted as a result of column saturation during chemistry, and therefore has not been included in subsequent analyses.

[d] Experiments *239* and *240* in BN capsules (all others in MgO capsules).

[e] Time and temperatures are those of equilibration (details of procedure in **Section 2.2**).

[f] Bulk silicate experiments to assess loss of isotopically light Zn.

Table 3 – Internal and external precision for complete isotopic methodology, and comparison to geostandards (MC-ICPMS)

| Standard | Sample | $\delta^{66}Zn$ | $2\sigma$ | $\delta^{68}Zn$ | $2\sigma$ | n |
|---|---|---|---|---|---|---|
| **BHVO-2** | BHVO-2 #1-1 | 0.32 | 0.02 | 0.61 | 0.05 | 4 |
| | BHVO-2 #1-2 | 0.29 | 0.02 | 0.59 | 0.04 | 4 |
| | BHVO-2 #1-3 | 0.29 | 0.04 | 0.56 | 0.03 | 3 |
| | *BHVO-2 #1 average* | 0.30 | 0.03 | 0.59 | 0.05 | |
| | BHVO-2 #2 | 0.30 | 0.04 | 0.64 | 0.06 | 4 |
| | ***BHVO-2 overall average*** | ***0.30*** | ***0.02*** | ***0.60*** | ***0.07*** | |
| **AGV-2** | AGV-2 #1 | 0.25 | 0.08 | 0.53 | 0.08 | 6 |
| **Zn dopant**[a] | Zn dopant | 0.13 | 0.01 | 0.25 | 0.04 | 5 |
| | Zn dopant #2[b] | 0.15 | 0.02 | 0.28 | 0.04 | 5 |
| | Zn dopant #3[c] | 0.15 | 0.05 | 0.30 | 0.11 | 11 |
| | Zn dopant #4[d] | 0.16 | 0.08 | 0.30 | 0.18 | 6 |
| | Zn dopant #5 | 0.11 | 0.08 | 0.21 | 0.17 | 5 |
| | ***Zn dopant overall average*** | ***0.14*** | ***0.04*** | ***0.27*** | ***0.08*** | |

[a] Zn dopant refers to the Zn powder used for all experiments (*Alfa Aesar*, -100 mesh, 99.9% metals basis).

[b] 'MORB (mix)' from Table 2

[c] 'Fe-FeS + MORB' from Table 2

[d] Separate Fe-FeS + MORB starting material for 'M' series experiments

Table 4 – ΔIW and log$K_e$ values of all experiments (from EPMA measurements)

| Starting Material | Experiment | Pressure Media | T (K) | t (min) | ΔIW | log$K_e$ | 2SE[a] |
|---|---|---|---|---|---|---|---|
| **Fe-FeS[b] + MORB** | *162* | Talc-Pyrex | 1673 | 5 | -1.71 | -1.45 | 0.17 |
| | *163* | Talc-Pyrex | 1673 | 15 | -1.80 | -1.49 | 0.19 |
| | *158* | Talc-Pyrex | 1673 | 30 | -1.55 | -1.60 | 0.03 |
| | *161* | Talc-Pyrex | 1673 | 60 | -1.80 | -1.71 | 0.11 |
| | *170* | BaCO$_3$ | 1673 | 120 | -1.78 | -1.62 | 0.04 |
| | *239[c]* | Talc-Pyrex | 1673 | 30 | -2.52 | -1.44 | 0.07 |
| | *240[c]* | Talc-Pyrex | 1673 | 60 | -2.47 | -1.55 | 0.07 |
| | *166* | Talc-Pyrex | 1823 | 15 | -1.89 | -1.21 | 0.25 |
| | *174* | BaCO$_3$ | 1973 | 3 | -1.88 | -1.20 | 0.25 |
| **Fe-FeS[b] + HPLG** | *192[d]* | Talc-Pyrex | 1473 | 60 | -2.45 | -1.55 | 0.18 |
| | *189[d]* | Talc-Pyrex | 1573 | 60 | -2.30 | -1.21 | 0.22 |
| | *181* | Talc-Pyrex | 1673 | 30 | -2.17 | -1.56 | 0.09 |
| **Fe + MORB** | *160* | BaCO$_3$ | 1973 | 5 | -2.38 | -1.16 | 0.05 |
| | *167* | BaCO$_3$ | 1973 | 3 | -2.33 | -1.19 | 0.06 |
| | *168* | BaCO$_3$ | 2123 | 1 | -2.68 | -1.12 | 0.13 |
| | *165* | BaCO$_3$ | 2273 | 1 | -2.36 | -0.87 | 0.07 |
| **Fe-Sn + MORB** | *171* | Talc-Pyrex | 1673 | 30 | -2.10 | -1.31 | 0.04 |
| | *173* | Talc-Pyrex | 1823 | 15 | -2.19 | -1.06 | 0.02 |
| | *172* | BaCO$_3$ | 1973 | 3 | -2.27 | -1.03 | 0.07 |

[a] Errors for log$K_e$ values have been propagated from EPMA measurements. Standard error (2SE) has been reported as S- and Sn-bearing metallic phases displayed quench textures (local heterogeneities) at the scale of the EPMA raster dimensions (20-30μm$^2$), leading to high standard deviations that reflect local heterogeneities and not instrument error. When this is the case, standard error more accurately reflects instrument error as it incorporates the number of measurements taken as well as the deviation among the measurements (after Chabot et al., 2009; Siebert et al., 2011).

[b] log$K_e$ values for all S-bearing experiments have been corrected for the influence of sulfur after Wood et al. (2014).

[c] Experiments *239* and *240* in BN capsules (all others in MgO capsules).

[d] Time and temperature are those of equilibration (details of procedure in **Section 2.2**).

**Figures**

Figure 1

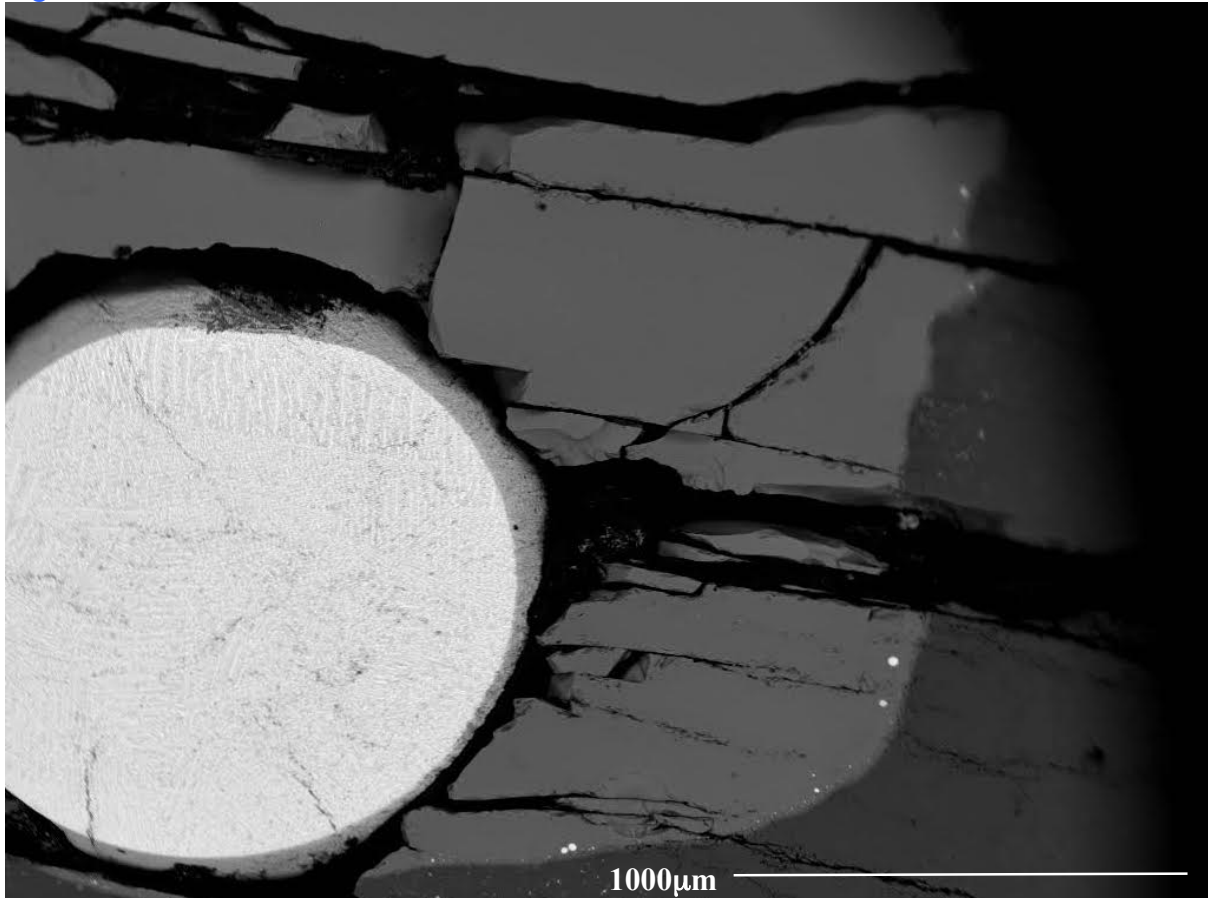

Figure 2

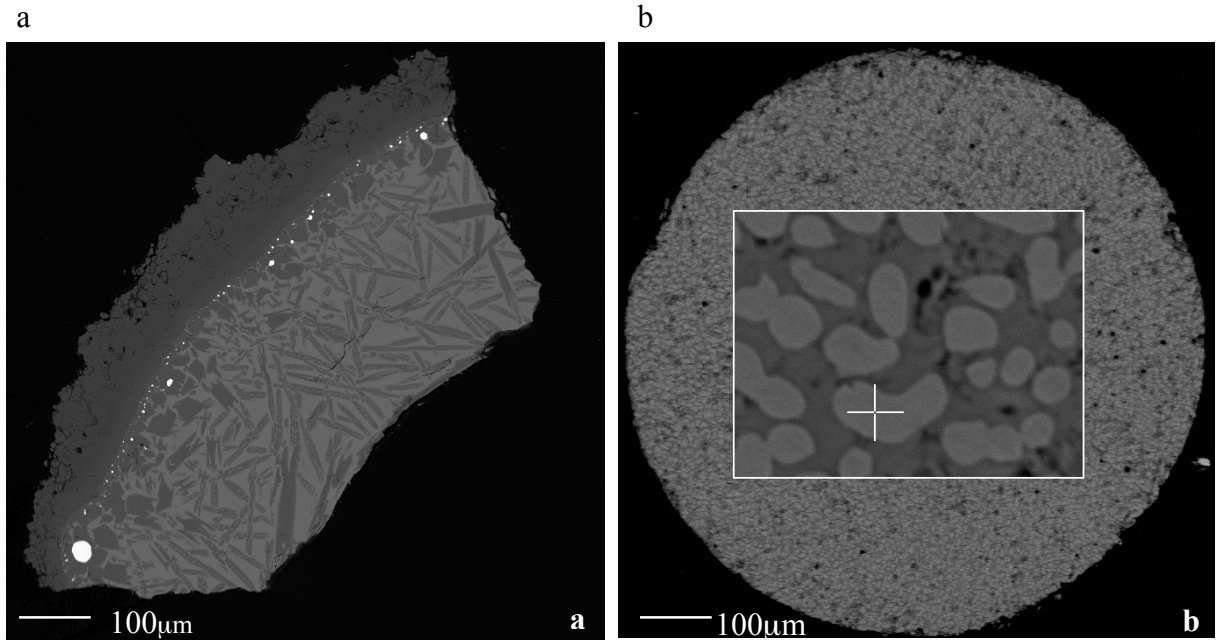

Figure 3

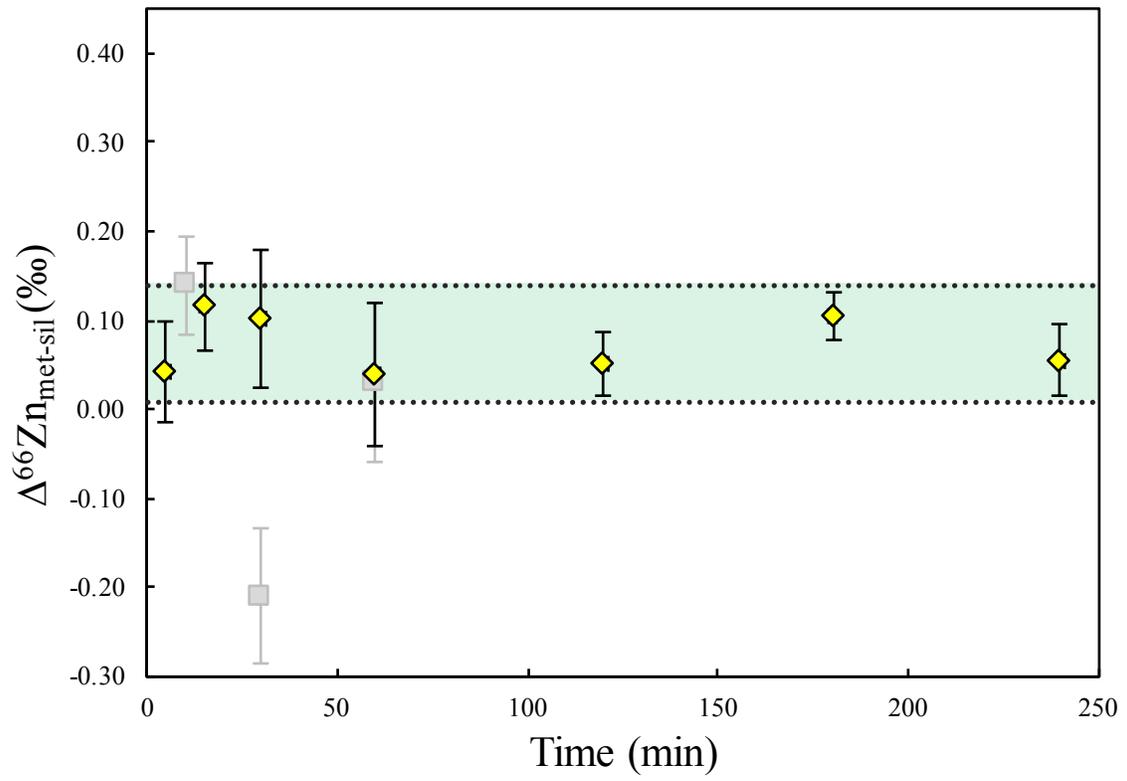

Figure 4

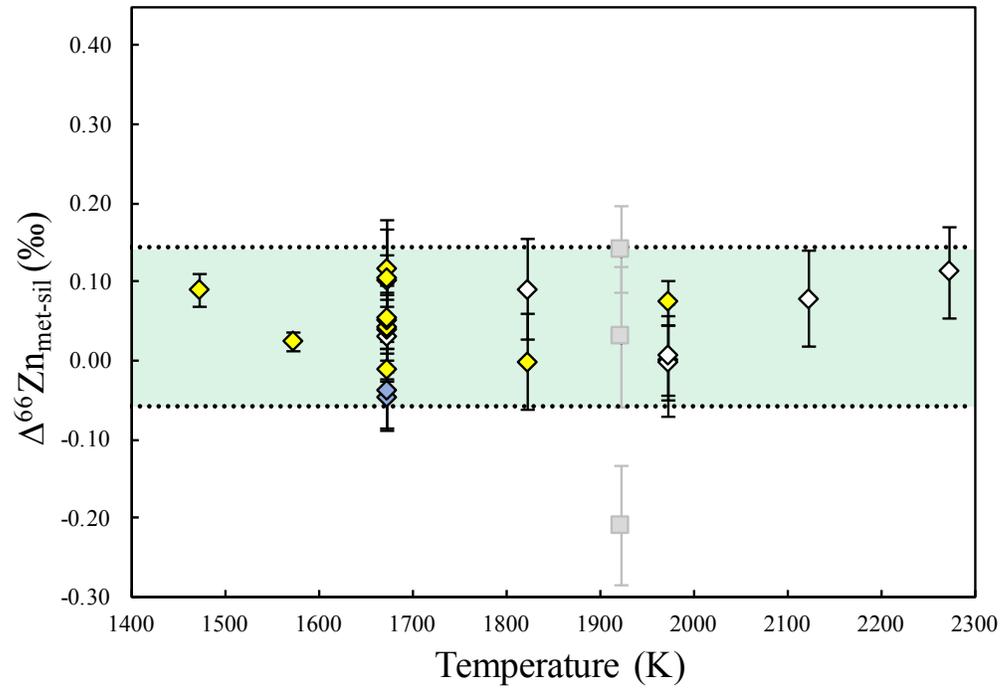

a)   b)

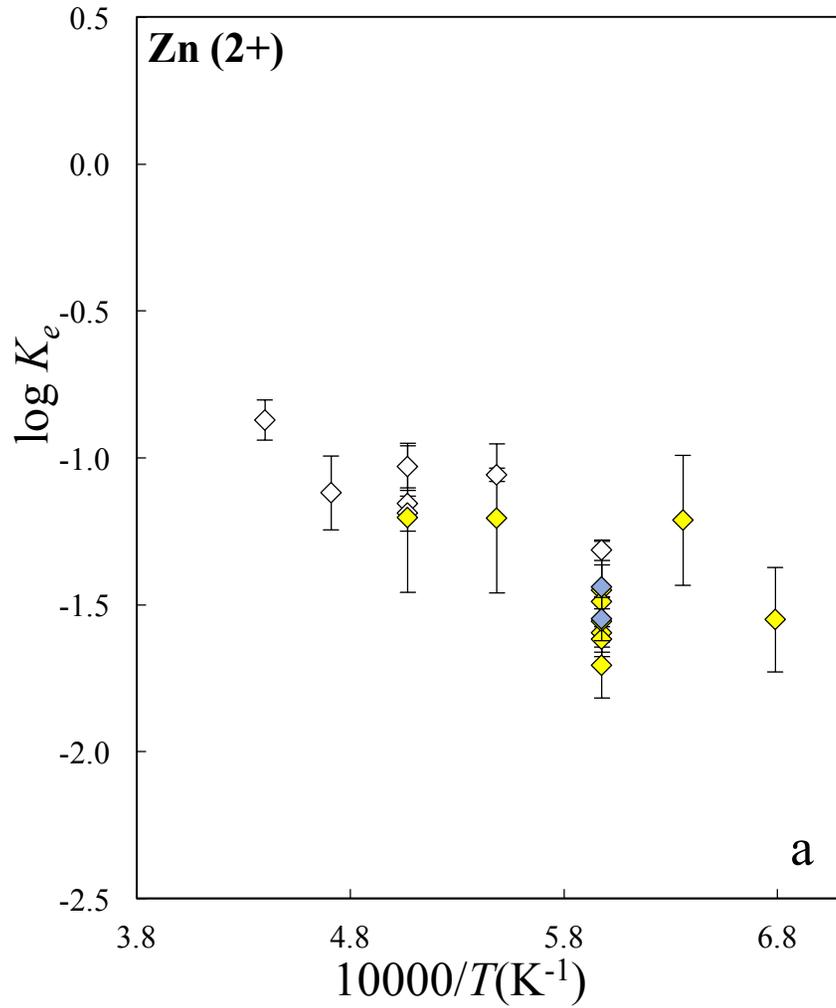
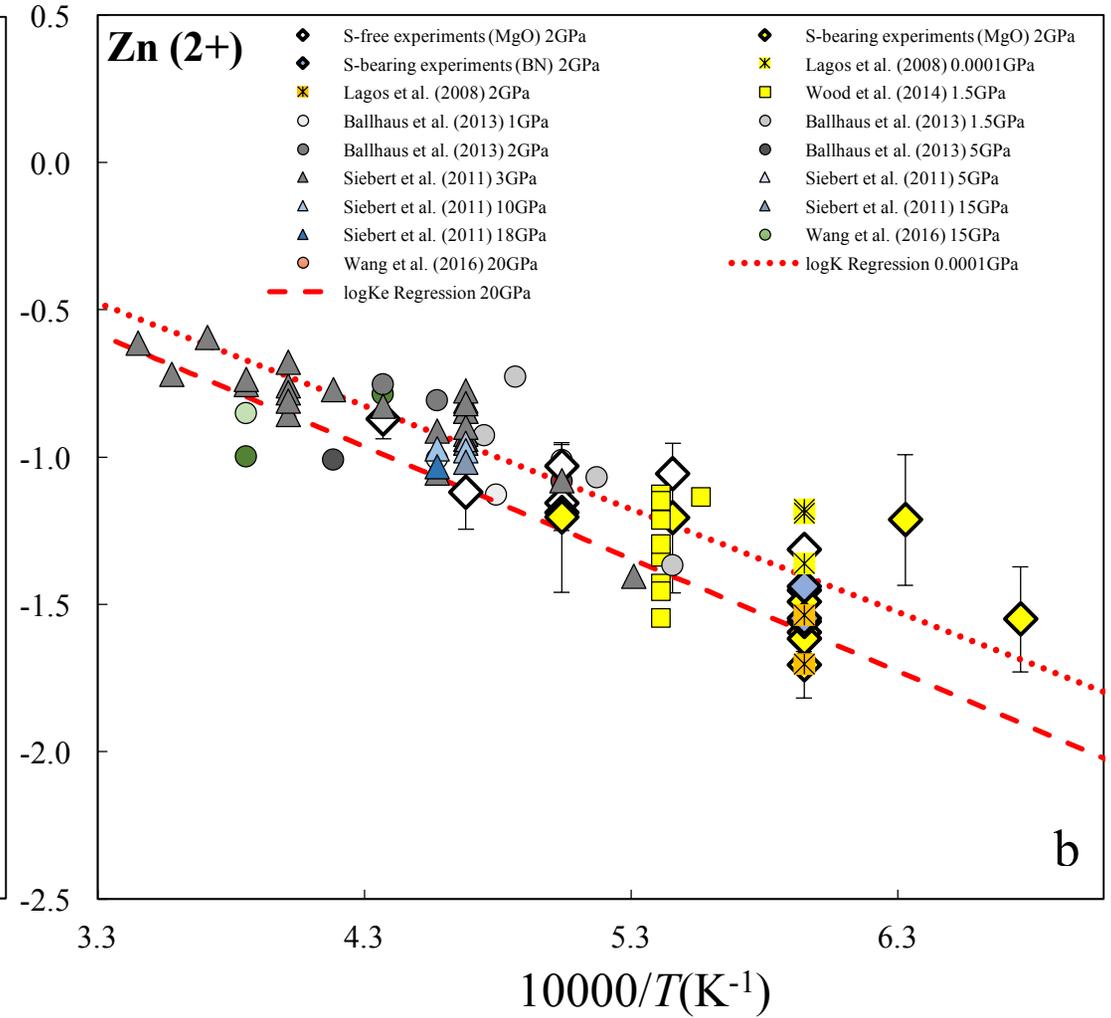

**Figure Captions**

Fig. 1: Backscatter electron image of a typical experimental charge. A large main metallic spherule (white/grey) surrounded by an a-crystalline glassy matrix (light grey), with a few stranded metallic blobs (bright spots) along the boundary with the capsule (dark grey).

Fig. 2: (**a**) BSE image of silicate phase at high temperature (exp. *167*, 1973K) with skeletal olivine (dark grey dendrites) and stranded metallic droplets (white spots) along the sample-capsule interface due to MgO enrichment at high temperatures. (**b**) BSE image of typical quench texture of S-bearing experiments (exp. *166* displayed), with an FeS-rich matrix (dark grey) containing homogeneously dispersed Fe-rich (light grey) globules that exsolved upon quenching. Zoomed rectangular area representative of a typical raster screen used for EPMA analysis (20-30μm$^2$), and crosshair gives representative scale of a point analysis (i.e. too small for integrative averaging).

Fig. 3: $\Delta^{66}Zn_{met-sil}$ as a function of time for Fe-FeS + MORB time series experiments (2 GPa, 1673K). Yellow diamonds indicate S-bearing experiments in MgO; grey squares indicate experiments from Bridgestock et al. (2014) in MgO for comparison. Errors for all $\Delta^{66}Zn_{met-sil}$ reported as one standard deviation (σ), with a shaded (green) ±σ envelope for data in the current study. No statistical trend of Zn metal-silicate isotope fractionation with time was observed.

Fig. 4: $\Delta^{66}Zn_{met-sil}$ as a function of temperature (at 2 GPa). Yellow diamonds indicate S-bearing experiments in MgO capsules; blue diamonds indicate S-bearing experiments in BN capsules; white diamonds indicate nominally S-free experiments in MgO capsules; grey squares indicate experiments from Bridgestock et al. (2014) in MgO capsules (1.5 GPa). Errors for all $\Delta^{66}Zn_{met-sil}$

reported as one standard deviation (σ), with a shaded (green) ±σ envelope for data in the current study. Zn metal-silicate isotope fractionation did not vary as a function of temperature.

Fig. 5: (**a**) Exchange coefficients plotted as a function of reciprocal temperature for all experiments (2 GPa). Yellow diamonds indicate S-bearing experiments in MgO capsules; blue diamonds indicate S-bearing experiments in BN capsules; white diamonds indicate nominally S-free experiments in MgO capsules. Experiments display a steady decrease in log$K_e$ as a function of reciprocal temperature, a trend in good agreement with literature data (Corgne et al., 2008; Siebert et al., 2011). Error (2SE) has been propagated from EPMA measurements. S-bearing experiments have been corrected for the effect of S on Zn metal-silicate partitioningar (Wood et al., 2014). (**b**) Data compilation with multilinear regression fit (see ***Section 4.3.1***). Data from this study (large diamonds in bold; same coloring as Fig. 5a) plotted with data from literature (Ballhaus et al., 2013; Lagos et al., 2008; Siebert et al., 2011; Wang et al., 2016; Wood et al., 2014). Icons for all references indicated in the legend. All S-bearing experiments indicated by yellow coloring; dark grey coloring indicates graphite capsules. Shading indicates pressure, where darker color qualitatively indicates higher pressure. Data shows a distinct effect of temperature, as well as a modest effect of pressure, on metal-silicate partitioning of Zn. All exchange coefficients have been calculated directly from concentration data. log$K_e$ regressions at .0001 GPa and 20 GPa have been calculated directly from our solution to Equation 8 (see ***Section 4.3.1*** for details).

# Appendices

Table A1 - Zn isotopic signature of all experimental runs by MC-ICPMS

| Starting Material | Experiment | δ$^{66}$Zn$_{met}$ | 2σ | δ$^{68}$Zn$_{met}$ | 2σ | n | δ$^{66}$Zn$_{sil}$ | 2σ | δ$^{68}$Zn$_{sil}$ | 2σ | n | Δ$^{66}$Zn$_{met-sil}$ | σ$^a$ | Δ$^{68}$Zn$_{met-sil}$ | σ$^a$ |
|---|---|---|---|---|---|---|---|---|---|---|---|---|---|---|---|
| **Fe-FeS + MORB** | 162 | 0.28 | 0.08 | 0.54 | 0.14 | 4 | 0.24 | 0.08 | 0.46 | 0.15 | 4 | 0.04 | 0.06 | 0.08 | 0.10 |
| | 163 | 0.28 | 0.04 | 0.54 | 0.09 | 3 | 0.17 | 0.09 | 0.33 | 0.16 | 8 | 0.12 | 0.05 | 0.22 | 0.09 |
| | 158 | 0.30 | 0.12 | 0.58 | 0.17 | 5 | 0.20 | 0.10 | 0.38 | 0.18 | 6 | 0.10 | 0.08 | 0.20 | 0.12 |
| | 161 | 0.28 | 0.14 | 0.53 | 0.29 | 5 | 0.25 | 0.08 | 0.46 | 0.17 | 4 | 0.04 | 0.08 | 0.07 | 0.17 |
| | M2 | 0.29 | 0.06 | 0.62 | 0.11 | 6 | 0.24 | 0.04 | 0.53 | 0.07 | 6 | 0.05 | 0.04 | 0.09 | 0.07 |
| | M1 | 0.43 | 0.05 | 1.07 | 0.11 | 6 | 0.32 | 0.02 | 0.75 | 0.05 | 6 | 0.10 | 0.03 | 0.32 | 0.06 |
| | M3 | 0.44 | 0.05 | 0.88 | 0.10 | 6 | 0.38 | 0.07 | 0.75 | 0.11 | 6 | 0.05 | 0.04 | 0.13 | 0.08 |
| | 170 | 0.65 | 0.07 | 1.27 | 0.13 | 8 | 0.16 | 0.06 | 0.31 | 0.10 | 4 | 0.49 | 0.05 | 0.96 | 0.08 |
| | 239 | 0.13 | 0.04 | 0.26 | 0.08 | 6 | 0.18 | 0.07 | 0.35 | 0.12 | 6 | -0.05 | 0.04 | -0.09 | 0.07 |
| | 240 | 0.15 | 0.09 | 0.28 | 0.15 | 5 | 0.19 | 0.05 | 0.38 | 0.11 | 5 | -0.04 | 0.05 | -0.09 | 0.09 |
| | 166 | 0.21 | 0.12 | 0.42 | 0.22 | 8 | 0.21 | 0.01 | 0.41 | 0.06 | 3 | 0.00 | 0.06 | 0.01 | 0.12 |
| | 174 | 0.25 | 0.05 | 0.46 | 0.10 | 4 | 0.17 | 0.02 | 0.30 | 0.03 | 3 | 0.07 | 0.03 | 0.16 | 0.05 |
| **Fe-FeS + HPLG** | 192 | 0.33 | 0.03 | 0.65 | 0.05 | 5 | 0.24 | 0.03 | 0.47 | 0.03 | 5 | 0.09 | 0.02 | 0.18 | 0.03 |
| | 189 | 0.31 | 0.01 | 0.62 | 0.02 | 5 | 0.29 | 0.02 | 0.58 | 0.04 | 5 | 0.02 | 0.01 | 0.04 | 0.02 |
| | 181 | 0.22 | 0.02 | 0.43 | 0.05 | 5 | 0.23 | 0.02 | 0.44 | 0.02 | 5 | -0.01 | 0.01 | 0.00 | 0.03 |
| **Fe + MORB** | 160 | 0.21 | 0.08 | 0.40 | 0.16 | 6 | 0.21 | 0.12 | 0.38 | 0.20 | 5 | 0.00 | 0.07 | 0.02 | 0.13 |
| | 167 | 0.20 | 0.09 | 0.37 | 0.18 | 6 | 0.20 | 0.02 | 0.38 | 0.05 | 3 | 0.00 | 0.05 | -0.01 | 0.09 |
| | 168 | 0.20 | 0.10 | 0.38 | 0.20 | 6 | 0.12 | 0.07 | 0.25 | 0.07 | 4 | 0.08 | 0.06 | 0.13 | 0.11 |
| | 165 | 0.25 | 0.09 | 0.49 | 0.19 | 6 | 0.14 | 0.08 | 0.27 | 0.14 | 6 | 0.11 | 0.06 | 0.21 | 0.12 |
| **Fe-Sn + MORB** | 171 | 0.24 | 0.09 | 0.47 | 0.14 | 4 | 0.22 | 0.07 | 0.41 | 0.12 | 4 | 0.03 | 0.05 | 0.06 | 0.09 |
| | 173 | 0.27 | 0.09 | 0.51 | 0.17 | 3 | 0.18 | 0.09 | 0.34 | 0.17 | 4 | 0.09 | 0.06 | 0.17 | 0.12 |
| | 172 | 0.20 | 0.07 | 0.38 | 0.14 | 4 | 0.19 | 0.07 | 0.38 | 0.14 | 4 | 0.01 | 0.05 | 0.00 | 0.10 |
| MORB (undoped) | 177 | --- | --- | --- | --- | | 0.42 | 0.04 | 0.84 | 0.07 | 5 | --- | --- | --- | --- |
| MORB | 179 | --- | --- | --- | --- | | 0.20 | 0.03 | 0.40 | 0.06 | 5 | --- | --- | --- | --- |
| MORB (undoped) | --- | --- | --- | --- | --- | | 0.34 | 0.03 | 0.71 | 0.06 | 5 | --- | --- | --- | --- |
| Zn dopant | Zn dopant | 0.13 | 0.01 | 0.25 | 0.04 | 5 | --- | --- | --- | --- | | --- | --- | --- | --- |
| MORB (mix) | Zn dopant #2 | --- | --- | --- | --- | | 0.15 | 0.02 | 0.28 | 0.04 | 5 | --- | --- | --- | --- |
| Fe-FeS + MORB | Zn dopant #3 | --- | --- | --- | --- | | 0.15 | 0.05 | 0.30 | 0.11 | 11 | --- | --- | --- | --- |
| Fe-FeS + MORB | Zn dopant #4 | --- | --- | --- | --- | | 0.16 | 0.08 | 0.30 | 0.18 | 6 | --- | --- | --- | --- |
| Zn dopant | Zn dopant #5 | 0.11 | 0.08 | 0.21 | 0.17 | 5 | --- | --- | --- | --- | | --- | --- | --- | --- |

$^a$ Standard deviation (1σ) is the conventional unit reported for Δ$^{66}$Zn$_{met-sil}$ (2σ conventionally reported for δ$^{66}$Zn).

Table A2 – Average major and minor element compositions of quenched metallic (wt%) and silicate (oxide wt%) melts by EPMA

| Experiment #[a] | 179[b] | 2SE | 162 | 2SE | 163 | 2SE | 158 | 2SE | 161 | 2SE | 170 | 2SE | 239 | 2SE |
|---|---|---|---|---|---|---|---|---|---|---|---|---|---|---|
| **Metal (wt%)** | | | | | | | | | | | | | | |
| O | --- | --- | 1.987 | 0.112 | 2.029 | 0.211 | 2.284 | 0.190 | 2.053 | 0.193 | 2.652 | 0.213 | 0.659 | 0.078 |
| S | --- | --- | 19.061 | 0.901 | 19.968 | 0.606 | 19.409 | 0.696 | 20.064 | 0.527 | 20.639 | 0.727 | 18.154 | 0.253 |
| Fe | --- | --- | 76.824 | 0.770 | 76.247 | 0.803 | 76.356 | 0.875 | 76.148 | 0.652 | 73.882 | 0.987 | 78.155 | 0.534 |
| Zn | --- | --- | 0.668 | 0.241 | 0.714 | 0.305 | 0.472 | 0.032 | 0.435 | 0.110 | 0.533 | 0.050 | 1.574 | 0.270 |
| Sn | --- | --- | --- | --- | --- | --- | --- | --- | --- | --- | --- | --- | --- | --- |
| P | --- | --- | 0.120 | 0.007 | 0.116 | 0.005 | 0.113 | 0.004 | 0.126 | 0.007 | 0.090 | 0.005 | --- | --- |
| **Total** | | | **98.66** | | **99.07** | | **98.63** | | **98.83** | | **97.79** | | **98.54** | |
| | | | | | | | | | | | | | | |
| **Silicate (oxide wt%)** | | | | | | | | | | | | | | |
| $Na_2O$ | 3.183 | 0.088 | 3.229 | 0.199 | 3.096 | 0.047 | 3.113 | 0.047 | 3.174 | 0.025 | 3.508 | 0.061 | 2.675 | 0.030 |
| MgO | 10.353 | 0.435 | 10.298 | 1.430 | 12.811 | 0.444 | 11.597 | 0.138 | 12.843 | 0.177 | 12.397 | 0.156 | 7.579 | 0.043 |
| $SiO_2$ | 47.037 | 0.314 | 43.148 | 0.685 | 44.550 | 0.295 | 42.246 | 0.500 | 44.034 | 0.444 | 43.730 | 0.259 | 48.157 | 0.206 |
| $Al_2O_3$ | 14.578 | 0.085 | 13.879 | 0.129 | 13.813 | 0.072 | 13.368 | 0.146 | 13.859 | 0.072 | 14.685 | 0.081 | 15.354 | 0.070 |
| $K_2O$ | 0.657 | 0.026 | 0.638 | 0.051 | 0.632 | 0.012 | 0.637 | 0.018 | 0.652 | 0.009 | 0.726 | 0.015 | 0.645 | 0.006 |
| CaO | 9.285 | 0.051 | 9.328 | 0.046 | 9.210 | 0.037 | 8.724 | 0.041 | 9.293 | 0.046 | 9.107 | 0.026 | 9.571 | 0.029 |
| $SO_2$ | 0.177 | 0.017 | 0.198 | 0.018 | 0.226 | 0.009 | 0.240 | 0.015 | 0.238 | 0.013 | 0.184 | 0.013 | 0.110 | 0.013 |
| FeO | 7.737 | 0.134 | 11.356 | 0.354 | 10.461 | 0.067 | 13.685 | 0.176 | 10.448 | 0.049 | 10.248 | 0.051 | 4.908 | 0.141 |
| $TiO_2$ | 1.572 | 0.027 | 1.581 | 0.037 | 1.535 | 0.031 | 1.448 | 0.021 | 1.584 | 0.033 | 1.708 | 0.037 | 1.722 | 0.009 |
| ZnO | 0.975 | 0.062 | 1.446 | 0.183 | 1.524 | 0.025 | 1.714 | 0.039 | 1.522 | 0.050 | 1.510 | 0.044 | 1.405 | 0.053 |
| SnO | --- | --- | --- | --- | --- | --- | --- | --- | --- | --- | --- | --- | --- | --- |
| $P_2O_5$ | 0.366 | 0.007 | --- | --- | --- | --- | 0.188 | 0.040 | --- | --- | --- | --- | 0.156 | 0.015 |
| MnO | 0.152 | 0.009 | --- | --- | --- | --- | 0.146 | 0.031 | --- | --- | --- | --- | --- | --- |
| $B_2O_3$ | --- | --- | --- | --- | --- | --- | --- | --- | --- | --- | --- | --- | 4.832 | 0.180 |
| **Total** | **96.07** | | **95.10** | | **97.86** | | **97.11** | | **97.65** | | **97.80** | | **97.11** | |

[a] Excluding *179*, order of experiments is the same as in Table 4.
[b] Mid ocean ridge basalt (MORB) doped with 1wt% Zn, used as baseline experimental silicate composition (2 GPa and 1673K in a Talc-Pyrex assembly; 30 minute run duration).

| Experiment # | 240 | 2SE | 166 | 2SE | 174 | 2SE | 192 | 2SE | 189 | 2SE | 181 | 2SE | 160 | 2SE |
|---|---|---|---|---|---|---|---|---|---|---|---|---|---|---|
| Metal (wt%) | | | | | | | | | | | | | | |
| O | 0.732 | 0.068 | 2.663 | 0.219 | 3.332 | 0.748 | 1.565 | 0.278 | 1.687 | 0.096 | 2.123 | 0.157 | 0.939 | 0.042 |
| S | 17.706 | 0.356 | 19.258 | 0.757 | 20.473 | 0.892 | 15.100 | 5.422 | 18.163 | 0.574 | 17.636 | 0.864 | 0.172 | 0.042 |
| Fe | 78.109 | 0.536 | 75.494 | 1.317 | 72.603 | 1.030 | 80.009 | 5.760 | 76.305 | 0.823 | 76.818 | 1.068 | 96.663 | 0.269 |
| Zn | 1.314 | 0.223 | 1.070 | 0.624 | 1.246 | 0.718 | 0.859 | 0.350 | 1.625 | 0.819 | 0.681 | 0.090 | 0.873 | 0.025 |
| Sn | --- | --- | --- | --- | --- | --- | --- | --- | --- | --- | --- | --- | --- | --- |
| P | --- | --- | 0.120 | 0.006 | 0.173 | 0.123 | -0.002 | 0.003 | -0.002 | 0.003 | 0.002 | 0.002 | 0.180 | 0.022 |
| Total | 97.86 | | 98.60 | | 97.83 | | 97.53 | | 97.78 | | 97.26 | | 98.83 | |
| | | | | | | | | | | | | | | |
| Silicate (oxide wt%) | | | | | | | | | | | | | | |
| $Na_2O$ | 2.481 | 0.030 | 2.918 | 0.034 | 3.604 | 0.532 | 4.460 | 0.114 | 4.522 | 0.089 | 4.319 | 0.065 | 3.071 | 0.610 |
| MgO | 7.795 | 0.052 | 20.831 | 0.228 | 23.409 | 3.407 | 2.208 | 1.280 | 1.664 | 0.943 | 1.654 | 0.666 | 26.216 | 4.530 |
| $SiO_2$ | 48.392 | 0.222 | 40.926 | 0.137 | 32.940 | 0.765 | 69.981 | 0.924 | 70.042 | 0.548 | 68.315 | 0.967 | 34.090 | 0.590 |
| $Al_2O_3$ | 15.532 | 0.068 | 13.397 | 0.249 | 14.786 | 1.117 | 10.170 | 0.213 | 10.191 | 0.150 | 10.094 | 0.319 | 13.957 | 1.909 |
| $K_2O$ | 0.621 | 0.008 | 0.606 | 0.012 | 0.723 | 0.100 | 5.947 | 0.196 | 6.093 | 0.145 | 5.743 | 0.110 | 0.687 | 0.131 |
| CaO | 9.574 | 0.029 | 8.664 | 0.092 | 10.105 | 1.266 | --- | --- | --- | --- | --- | --- | 9.482 | 1.648 |
| $SO_2$ | 0.095 | 0.011 | 0.353 | 0.012 | 0.567 | 0.062 | 0.123 | 0.067 | 0.083 | 0.012 | 0.084 | 0.030 | 0.025 | 0.006 |
| FeO | 5.300 | 0.116 | 9.828 | 0.044 | 9.514 | 0.417 | 5.567 | 0.284 | 6.030 | 0.267 | 6.888 | 0.752 | 8.534 | 0.491 |
| $TiO_2$ | 1.736 | 0.009 | 1.422 | 0.027 | 1.551 | 0.148 | --- | --- | --- | --- | --- | --- | 1.577 | 0.250 |
| ZnO | 1.655 | 0.055 | 1.221 | 0.032 | 1.453 | 0.091 | 1.156 | 0.108 | 1.059 | 0.053 | 1.187 | 0.115 | 1.066 | 0.091 |
| SnO | --- | --- | --- | --- | --- | --- | --- | --- | --- | --- | --- | --- | --- | --- |
| $P_2O_5$ | 0.143 | 0.014 | 0.128 | 0.006 | 0.261 | 0.039 | --- | --- | --- | --- | --- | --- | --- | --- |
| MnO | --- | --- | 0.137 | 0.007 | 0.136 | 0.008 | 0.021 | 0.005 | 0.016 | 0.007 | 0.016 | 0.007 | --- | --- |
| $B_2O_3$ | 4.831 | 0.226 | --- | --- | --- | --- | --- | --- | --- | --- | --- | --- | --- | --- |
| Total | 98.16 | | 100.43 | | 99.05 | | 99.63 | | 99.70 | | 98.30 | | 98.70 | |

Table A2 (*continued*)

| Experiment # | 167 | 2SE | 168 | 2SE | 165 | 2SE | 171 | 2SE | 173 | 2SE | 172 | 2SE |
|---|---|---|---|---|---|---|---|---|---|---|---|---|
| Metal (wt%) | | | | | | | | | | | | |
| O | 1.091 | 0.061 | 1.103 | 0.146 | 1.260 | 0.412 | 1.289 | 0.052 | 1.458 | 0.073 | 1.364 | 0.042 |
| S | 0.158 | 0.030 | 0.193 | 0.139 | 0.234 | 0.122 | 0.178 | 0.018 | 0.215 | 0.019 | 0.216 | 0.027 |
| Fe | 96.068 | 0.367 | 95.610 | 0.402 | 95.049 | 0.432 | 70.394 | 0.309 | 67.943 | 0.522 | 71.949 | 0.732 |
| Zn | 0.796 | 0.038 | 1.028 | 0.092 | 1.019 | 0.145 | 0.560 | 0.041 | 0.940 | 0.046 | 0.987 | 0.054 |
| Sn | --- | --- | --- | --- | --- | --- | 26.269 | 0.220 | 28.125 | 0.469 | 22.988 | 0.727 |
| P | 0.180 | 0.014 | 0.161 | 0.052 | 0.227 | 0.094 | 0.168 | 0.006 | 0.189 | 0.009 | 0.168 | 0.010 |
| Total | 98.29 | | 98.10 | | 97.79 | | 98.86 | | 98.87 | | 97.67 | |
| | | | | | | | | | | | | |
| Silicate (oxide wt%) | | | | | | | | | | | | |
| $Na_2O$ | 2.904 | 0.608 | 1.293 | 0.491 | 1.703 | 0.474 | 3.148 | 0.023 | 2.832 | 0.026 | 2.992 | 0.856 |
| MgO | 25.050 | 5.234 | 40.577 | 3.894 | 38.309 | 3.830 | 13.687 | 0.291 | 20.681 | 0.200 | 28.276 | 5.884 |
| $SiO_2$ | 35.429 | 0.588 | 31.750 | 1.840 | 30.602 | 0.890 | 43.033 | 0.401 | 40.184 | 0.082 | 33.686 | 1.331 |
| $Al_2O_3$ | 13.389 | 1.802 | 10.445 | 2.240 | 8.852 | 1.845 | 13.875 | 0.107 | 13.216 | 0.225 | 12.524 | 2.376 |
| $K_2O$ | 0.595 | 0.088 | 0.321 | 0.128 | 0.410 | 0.109 | 0.638 | 0.009 | 0.566 | 0.011 | 0.656 | 0.159 |
| CaO | 8.896 | 1.727 | 4.892 | 1.586 | 5.903 | 1.174 | 9.332 | 0.046 | 8.345 | 0.075 | 9.146 | 2.180 |
| $SO_2$ | 0.027 | 0.006 | 0.028 | 0.007 | 0.028 | 0.016 | 0.021 | 0.007 | 0.023 | 0.012 | 0.027 | 0.011 |
| FeO | 8.857 | 0.807 | 6.198 | 0.902 | 8.690 | 0.239 | 9.104 | 0.048 | 8.245 | 0.038 | 8.171 | 0.862 |
| $TiO_2$ | 1.519 | 0.246 | 0.933 | 0.227 | 0.945 | 0.157 | 1.520 | 0.044 | 1.425 | 0.033 | 1.465 | 0.303 |
| ZnO | 1.098 | 0.098 | 0.759 | 0.137 | 0.667 | 0.059 | 1.447 | 0.049 | 1.260 | 0.027 | 1.164 | 0.136 |
| SnO | --- | --- | --- | --- | --- | --- | 0.259 | 0.011 | 0.256 | 0.010 | 0.279 | 0.084 |
| $P_2O_5$ | --- | --- | --- | --- | --- | --- | --- | --- | 0.158 | 0.008 | --- | --- |
| MnO | --- | --- | --- | --- | --- | --- | --- | --- | 0.130 | 0.010 | --- | --- |
| $B_2O_3$ | --- | --- | --- | --- | --- | --- | --- | --- | --- | --- | --- | --- |
| Total | 97.76 | | 97.20 | | 96.11 | | 96.06 | | 97.32 | | 98.39 | |

Table A3 – Partitioning data for parameterization of Zn metal-silicate partitioning via least squares multi-variable regression ($N_{total}$ = 78) (see **Section 4.3.1** and Equation 8).

| Source | Experiment | log$K_e$[a] | T (K) | P (GPa) |
|---|---|---|---|---|
| *This study (N = 19)* | *158* | -1.60 | 1673 | 2.0 |
| | *160* | -1.16 | 1973 | 2.0 |
| | *161* | -1.71 | 1673 | 2.0 |
| | *162* | -1.45 | 1673 | 2.0 |
| | *163* | -1.49 | 1673 | 2.0 |
| | *165* | -0.87 | 2273 | 2.0 |
| | *166* | -1.21 | 1823 | 2.0 |
| | *167* | -1.19 | 1973 | 2.0 |
| | *168* | -1.12 | 2123 | 2.0 |
| | *170* | -1.62 | 1673 | 2.0 |
| | *171* | -1.32 | 1673 | 2.0 |
| | *172* | -1.03 | 1973 | 2.0 |
| | *173* | -1.06 | 1823 | 2.0 |
| | *174* | -1.20 | 1973 | 2.0 |
| | *181* | -1.56 | 1673 | 2.0 |
| | *189* | -1.21 | 1573 | 2.0 |
| | *192* | -1.55 | 1473 | 2.0 |
| | *239* | -1.44 | 1673 | 2.0 |
| | *240* | -1.55 | 1673 | 2.0 |
| *Lagos et al., 2008 (N = 5)* | *PC429* | -1.70 | 1673 | 2.0 |
| | *PC430* | -1.54 | 1673 | 2.0 |
| | *CAP1* | -1.18 | 1673 | 0.0001 |
| | *CAP2* | -1.19 | 1673 | 0.0001 |
| | *CAP4* | -1.36 | 1673 | 0.0001 |
| *Siebert et al., 2011 (N = 31)* | *94* | -1.05 | 2173 | 3.0 |
| | *96* | -0.76 | 2473 | 3.0 |
| | *97* | -0.75 | 2573 | 3.0 |
| | *116* | -0.92 | 2123 | 3.0 |
| | *118* | -0.91 | 2173 | 3.0 |
| | *120* | -0.83 | 2273 | 3.0 |
| | *121* | -0.77 | 2373 | 3.0 |
| | *122* | -0.85 | 2473 | 3.0 |
| | *124* | -0.81 | 2123 | 3.0 |
| | *126* | -0.84 | 2123 | 3.0 |

[a] Where applicable, log$K_e$ values have been corrected for the influence of C (after Siebert et al., 2011) and S (after Wood et al., 2014)

| Source | Experiment | log$K_e^a$ | T (K) | P (GPa) |
|---|---|---|---|---|
| *Siebert et al., 2011* | *127* | -1.08 | 1973 | 3.0 |
| | *128* | -1.40 | 1873 | 3.0 |
| | *129* | -0.74 | 2573 | 3.0 |
| | *130* | -0.59 | 2673 | 3.0 |
| | *131* | -0.72 | 2773 | 3.0 |
| | *132* | -0.61 | 2873 | 3.0 |
| | *133* | -0.68 | 2473 | 3.0 |
| | *134* | -0.78 | 2473 | 3.0 |
| | *135* | -0.81 | 2473 | 3.0 |
| | *139* | -0.78 | 2123 | 3.0 |
| | *140* | -0.94 | 2123 | 3.0 |
| | *141* | -0.84 | 2123 | 2.0 |
| | *142* | -0.95 | 2123 | 1.0 |
| | *143* | -0.90 | 2123 | 3.0 |
| | *144* | -0.82 | 2123 | 3.0 |
| | *145* | -0.83 | 2123 | 0.5 |
| | *MA17* | -0.97 | 2173 | 10 |
| | *MA18* | -0.98 | 2123 | 10 |
| | *MA20* | -0.95 | 2123 | 5 |
| | *MA21* | -1.02 | 2123 | 15 |
| | *MA22* | -1.03 | 2173 | 18 |
| *Ballhaus et al., 2013 (N = 11)* | *A9* | -1.01 | 2173 | 1.0 |
| | *A11* | -1.08 | 1973 | 5.0 |
| | *A12* | -1.01 | 2373 | 5.0 |
| | *A15* | -0.86 | 2173 | 2.0 |
| | *A17* | -0.62 | 2273 | 2.0 |
| | *A20* | -1.01 | 1973 | 1.0 |
| | *A21* | -1.13 | 2073 | 1.0 |
| | *A-40* | -1.37 | 1823 | 1.5 |
| | *A-41* | -1.07 | 1923 | 1.5 |
| | *A-42* | -0.73 | 2043 | 1.5 |
| | *A-43* | -0.93 | 2093 | 1.5 |
| *Wood et al., 2014 (N = 9)* | *KK18-2* | -1.13 | 1838 | 1.5 |
| | *KK19-1* | -1.14 | 1788 | 1.5 |

| Source | Experiment | log$K_e^a$ | T (K) | P (GPa) |
|---|---|---|---|---|
| *Wood et al., 2014* | *KK19-2* | -1.15 | 1838 | 1.5 |
| | *KK20-1* | -1.43 | 1838 | 1.5 |
| | *KK20-2* | -1.55 | 1838 | 1.5 |
| | *KK21-1* | -1.21 | 1838 | 1.5 |
| | *KK22-1* | -1.34 | 1838 | 1.5 |
| | *KK23-1* | -1.30 | 1838 | 1.5 |
| | *KK23-2* | -1.46 | 1838 | 1.5 |
| *Wang et al., 2016 (N = 3)* | *Z1200b* | -0.79 | 2273 | 20 |
| | *Z1201b* | -0.85 | 2573 | 15 |
| | *Z1206b* | -1.00 | 2573 | 20 |